\documentclass[a4paper, superscriptaddress, twocolumn, prl, showpacs, longbibliography]{revtex4-1}

\usepackage{dcolumn}
\usepackage{bm}
\usepackage[colorlinks=true,citecolor=blue,linkcolor=blue,pdfhighlight =/O]{hyperref}
\usepackage{graphicx}
\usepackage{amsmath}
\usepackage{dsfont}
\usepackage{mathtools}
\usepackage[dvipsnames]{xcolor}
\usepackage{soul}
\usepackage[normalem]{ulem}

\newcommand{\mr}[1]{\mathrm{#1}}
\newcommand{\be}{\begin{equation}}
\newcommand{\ee}{\end{equation}}

\newcommand{\kb}{k_{\mr{B}}}

\newcommand{\mk}{\;\mr{mK}}

\newcommand{\mv}{\;\mr{mV}}

\newcommand{\mev}{\;\mr{meV}}

\newcommand{\muv}{\;\mu\mr{V}}
\newcommand{\muev}{\;\mu\mr{eV}}

\newcommand{\ethls}{\epsilon_{\mr{th}}}

\newcommand{\te}{T_{\mr{e}}}

\newcommand{\qd}{\mr{QD}}
\newcommand{\sns}{\mr{SNS}}

\newcommand{\ec}{E_{\mr{c}}}

\newcommand{\qdot}{\dot{Q}}
\newcommand{\qdotH}{\dot{Q}_{\mr{H}}}

\newcommand{\tb}{T_{\mr{b}}}

\newcommand{\vg}{V_{\mr{g}}}
\newcommand{\vb}{V_{\mr{b}}}

\newcommand{\ic}{I_{\mr{C}}}

\hypersetup{urlcolor=blue}

\begin{document}
\title{A Single-Quantum-Dot Heat Valve}
\author{B. Dutta}
\affiliation{\mbox{Univ.} Grenoble Alpes, CNRS, Grenoble INP, Institut N\' eel, 25 Avenue des Martyrs, 38042 Grenoble, France}
\author{D. Majidi}
\affiliation{\mbox{Univ.} Grenoble Alpes, CNRS, Grenoble INP, Institut N\' eel, 25 Avenue des Martyrs, 38042 Grenoble, France}
\author{N. W. Talarico}
\affiliation{QTF Centre of Excellence, Turku Centre for Quantum Physics, Department of Physics and Astronomy, University of Turku, 20014 Turku, Finland}
\author{N. Lo Gullo}
\affiliation{QTF Centre of Excellence, Turku Centre for Quantum Physics, Department of Physics and Astronomy, University of Turku, 20014 Turku, Finland}
\author{C. B. Winkelmann}
\affiliation{\mbox{Univ.} Grenoble Alpes, CNRS, Grenoble INP, Institut N\' eel, 25 Avenue des Martyrs, 38042 Grenoble, France}
\author{H. Courtois}
\affiliation{\mbox{Univ.} Grenoble Alpes, CNRS, Grenoble INP, Institut N\' eel, 25 Avenue des Martyrs, 38042 Grenoble, France}
\date{\today}

\begin{abstract}
We demonstrate gate control of electronic heat flow in a thermally-biased single-quantum-dot junction. Electron temperature maps taken in the immediate vicinity of the junction, as a function of the gate and bias voltages applied to the device, reveal clearly defined Coulomb diamond patterns revealing a maximum heat transfer right at the charge degeneracy point. The non-trivial bias and gate dependence of this heat valve results from both the quantum nature of the dot at the heart of device and its strong coupling to leads.
\end{abstract}

\maketitle

In the emerging field of quantum thermodynamics, heat transport and dissipation in a quantum electronic device is a fundamentally important topic \cite{Pop2010,Lee2013,Zotti2014,Dubi2011}. Gate-tunable single-quantum dot junctions \cite{Kouwenhoven2001} are paradigmatic test benches for quantum transport. Whereas the study of charge transport therein has already allowed the exploration of a large palette of physical effects at play, heat transport and thermoelectric properties have been investigated in a limited number of cases, \mbox{e.g.} in quantum dots formed in a two-dimensional electron gas (2DEG) \cite{Scheibner2005,Dzurak1997,Godijn1999} and in semiconducting nanowires \cite{Josefsson2018,Jaliel-PRL19,Prete-NanoLett19}. As opposed to charge transport processes, the understanding of electronic heat transport and generation across a nano-scale object is, experimentally, still in its infancy \cite{Kim2001,Huang2007,Tsutsui2008}. Local thermometry has been achieved only in a very limited number of quantum devices. The temperature dependence of the critical current of a superconducting weak link was used in scanning probe experiments to reveal for instance the scattering sites in high-mobility graphene \cite{Halbertal2016, Marguerite2019}. Yet, to date, these experiments are limited to temperatures above 1 K. At milliKelvin temperatures, local thermometry can be performed in quantum devices formed in a 2DEG by a variety of methods \cite{Molenkamp1992,Prance-PRL09} that have recently been pushed to quantitative accuracy \cite{Jezouin2013,Sivre2018,Banerjee2018}. Noise thermometry was applied to thermoelectric measurements in InAs nanowires \cite{Tikhonov2016}. In metallic devices, electronic thermometry is usually based on the temperature dependence of charge transport in superconducting hybrids, either in the tunnelling regime for Normal metal-Insulator-Superconductor (NIS) junctions \cite{Giazotto-RMP,Nahum1995} or at higher transparencies allowing for superconducting correlations \cite{Meschke-JLTP09,WangAPL18,Giazotto-Nature12}. This has recently allowed the realization of a photonic heat valve with a superconducting qubit coupled to heat reservoirs (probed by NIS probes) through coplanar waveguide resonators \cite{Ronzani-NatPhys18}.

The single electron transistor (SET) is an essential brick for the emerging field of quantum caloritronics \cite{Saira2007}. Building on the NIS thermometry technique, the thermal conductance of a metallic SET was measured \cite{Dutta-PRL17}. Despite the continuous density of states in the metallic island, electron interactions readily lead to striking deviations from the Wiedemann-Franz law \cite{Kubala-PRL08}. Going beyond this simple case, two questions arise: (i) how does such a SET behave thermally beyond equilibrium, that is, at finite voltage bias and/or at large temperature difference where both Joule heat and heat transport are to be taken into account, and (ii), if the central island is replaced by a quantum dot ($\qd$), how would the discrete nature of its energy spectrum manifest in the thermal properties of the device? In the weak coupling regime, the discreteness of the QD energy spectrum makes electronic transport processes strongly selective in energy. At zero net particle current, whatever the gate voltage, the heat flow is zero since electrons tunnel back and forth exactly at the energy level defined by the QD. The heat conductance is thus zero at all gate voltages. Heat transfer is predicted only at non-zero particle current, when the QD energy level is positioned just above or below the Fermi level of the hot lead, so that high-energy electrons can escape through the dot, or low-energy electrons can be injected there \cite{Edwards-APL93,Prance-PRL09}.

In this Letter, we report on the operation of a single-quantum-dot heat valve. The methodological novelty is to introduce local thermometry in a metallic single quantum dot device. When current flows through it, Joule dissipation results in a temperature increase following the usual Coulomb diamonds' pattern. At zero net charge current and charge degeneracy, the observed electronic heat transfer is the result of energy quantization in the dot combined with strong tunnel coupling to the leads.

\begin{figure}[!t]
	\includegraphics[width=0.99\columnwidth]{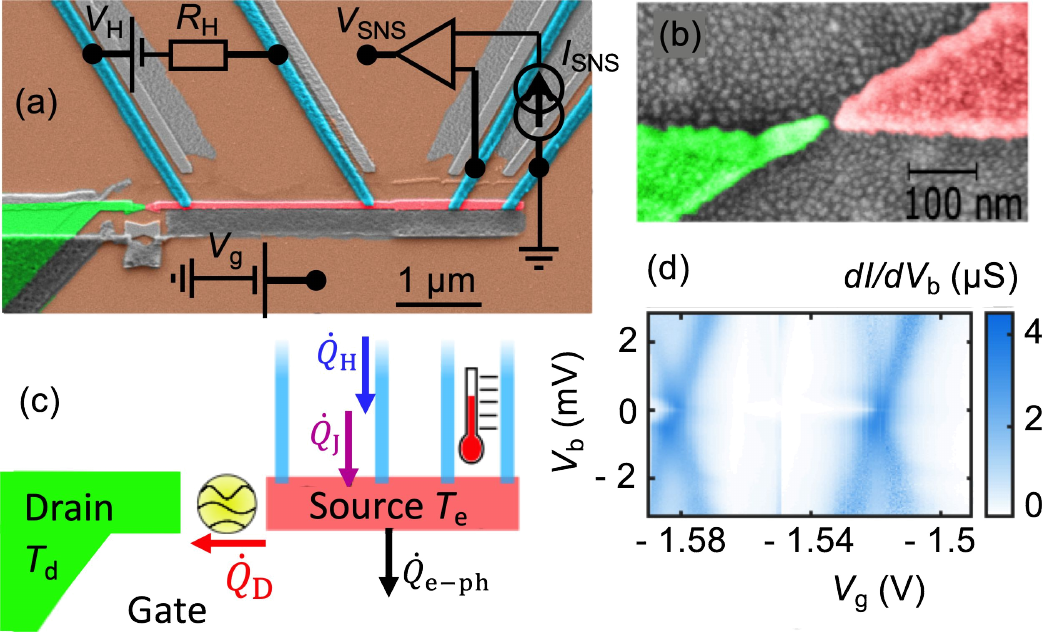}
	\caption{(a) False-colored SEM image of a typical device. The source is colored in red, the drain in green and the superconducting leads in blue. The circuit diagram shows the heat transport set-up. The longer (2.5 $\mu$m) $\sns$ junction is used as a heater driven by a constant \mbox{d.c.} battery and the shorter (700 nm) $\sns$ junction is used as a thermometer. (b) Zoomed-in view of the nano-gap between the $source$ and $drain$ created by electromigration and the nano-particles made by Au evaporation. (c) Schematic of the device, with the different heat flows to/from the source. (d) Differential conductance map of the device measured at 70 mK against the drain-source bias voltage $V_{\rm b}$ and the gate voltage $V_{\rm g}$ with no additional heating applied.}
	\label{fig:figure1}
\end{figure}

Figures \ref{fig:figure1}(a,b) display a colored scanning electron micrograph of a typical device as the one reported here, whereas \mbox{Fig. \ref{fig:figure1}(c)} shows a thermal diagram of the same, with the corresponding color code for each device element. 
The device is different from that in \mbox{Ref. \cite{Dutta-Nanolett19}} but it has the same geometry and the fabrication procedure is similar. The fabrication of the main part of the device is based on e-beam lithography, three-angle, Au thin film evaporation and lift-off. After the lift-off, we deposit a 1-2 nm thin Au layer on top of the whole device. Due to surface tension forces, this small amount of deposit leads to the formation of 5--10 nm diameter Au nanoparticles on the sample.
A bow-tie shaped Pt electromigration junction forms the central part of the device on which the Au nanoparticles form a dense layer of QDs, see \mbox{Fig. \ref{fig:figure1}(b)}. Here we have chosen Pt as the electromigrated material in order to ensure the source local density of states at the QD contact to be free of superconducting correlations induced by the nearest Al lead \cite{Kontos2004}.

The electromigration junction is connected on one side to a bulky {\it drain} electrode made of Au, in fairly good contact to the thermal bath at a temperature $\tb$, and on the other side to a narrow {\it source} electrode, again made of Au \cite{Dutta-Nanolett19}. Four Al leads provide contacts to the source through a transparent interface. At temperatures well below Al superconducting critical temperature, these leads are thermally insulating. The source is therefore fairly thermally decoupled from its environment. In the standard hot electron assumption, electron-electron equilibration is much faster than any other dynamical process. The source electrons are thus in a quasi-equilibrium state described by a Fermi-Dirac distribution at a temperature $\te$ that can significantly differ from $\tb$. The pair of closely spaced Al leads connected to the source forms an $\sns$ junction with a temperature-dependent critical current that will be used as an electronic thermometer. Conversely, the widely-spaced pair of Al leads forms instead a junction with a vanishing critical current, which allows it to be used as an ohmic heater. In contrast to prior work \cite{Dutta-PRL17}, we have chosen here transparent rather than tunnel contacts to the source, for two reasons. First, SNS junctions can provide less invasive thermometers than NIS junctions that are biased at a voltage of about $\Delta/e$, which in turn can lead to significant heating and cooling effects \cite{Giazotto-RMP}. Second, electromigration requires low access resistances, which is inherently incompatible with tunnel contacts. 

\begin{figure}[t]
	\includegraphics[width=0.99\columnwidth]{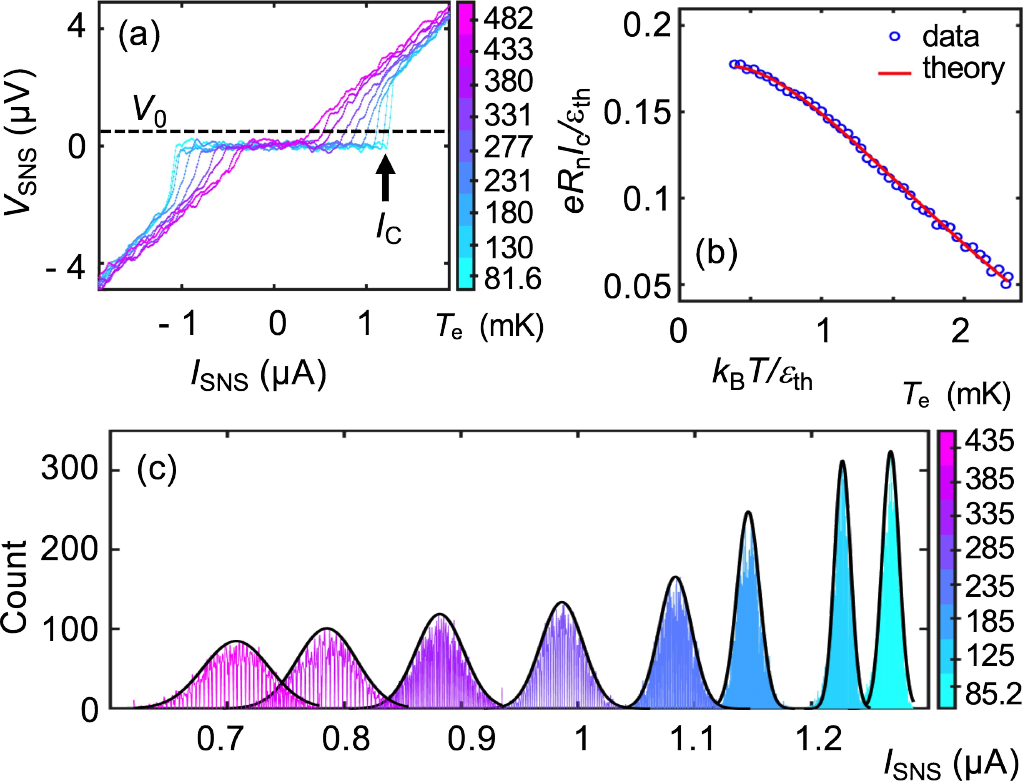}
	\caption{(a) \mbox{d.c.} IV characteristics of the $\sns$ thermometer junction at different bath temperature $\tb$, the current bias value at which the voltage exceeds a threshold $V_{\rm 0} \simeq 0.5 \, \mu$V defining the switching current. (b) The critical current $\ic$ as a function of the bath temperature, the axes being normalized. It is defined as the most probable switching current extracted from the histograms. The calibration curve (red solid line) is a fit with the theory \cite{Dubos-PRB01}. (c) Histogram of the stochastic switching current of the $\sns$ junction at different bath temperatures, with a fitted gaussian envelope for each.}
	\label{fig:figure2}
\end{figure}

The nanometer-sized gap was created within the Pt constriction by means of electromigration at a temperature of 4 K \cite{Park-APL99,Dutta-Nanolett19}. The device was further cooled {\it in situ} down to the cryostat base temperature of about 70 mK. Figure \ref{fig:figure1}(d) shows a differential conductance map of the $\qd$ junction as a function of bias and gate voltages, $V_{\rm b}$ and $V_{\rm g}$ respectively, with no additional heating. From the observed Coulomb diamonds, one finds a charging energy $\ec =$ 4 meV \cite{ChargingEnergy}. Our detailed analysis \cite{suppmat} provides a tunnel coupling $\hbar\Gamma$ value in the range 0.2 - 1.5 $\mev$, depending on the considered single energy level involved in low-bias electron transport at a given charge degeneracy point.
In spite of the large tunnel coupling $\hbar\Gamma\gg\kb T$, it is still not strong enough to induce Kondo effect.

We now move to the description of the electron thermometers. The critical current $\ic$ of an SNS junction is highly sensitive to the electronic temperature $\te$ in N. The relevant energy scale is the Thouless energy $\ethls=\hbar D/L^2$, where $D$ is the diffusion constant in N and $L$ is the junction length \cite{Dubos-PRB01}. For $\te> \ethls / k_{\rm B}$, $\ic$ decreases rapidly with increasing temperature, allowing it to be used as a secondary electron thermometer \cite{Meschke-JLTP09,WangAPL18}. In a single IV characteristic, the switching current is defined as the value of the current at which the voltage is larger than a threshold voltage chosen above the measurement noise level. \mbox{Figure \ref{fig:figure2}(a)} shows a series of such characteristics at different bath temperatures. Switching current histograms, together with a gaussian fit of their envelope, are shown in \mbox{Fig. \ref{fig:figure2}(c)} for a series of bath temperature values. The histogram width increases with the temperature, consistently with a Josephson energy fluctuating by $2 k_{\rm B} T$. In \mbox{Fig. \ref{fig:figure2}(b)}, the variation of the critical current with the bath temperature fits nicely the theoretical expectation \cite{Dubos-PRB01}, the latter being used as the thermometer calibration. The low Thouless energy \mbox{$\ethls$ $\sim$ 5 $\mu$eV} was chosen in order to avoid a saturation of $\ic$. The thermometer thus remains sensitive at low temperature, where thermal transport through the $\qd$ gains importance compared to other heat relaxation processes.

\begin{figure}[!t]
	\includegraphics[width=0.99\columnwidth]{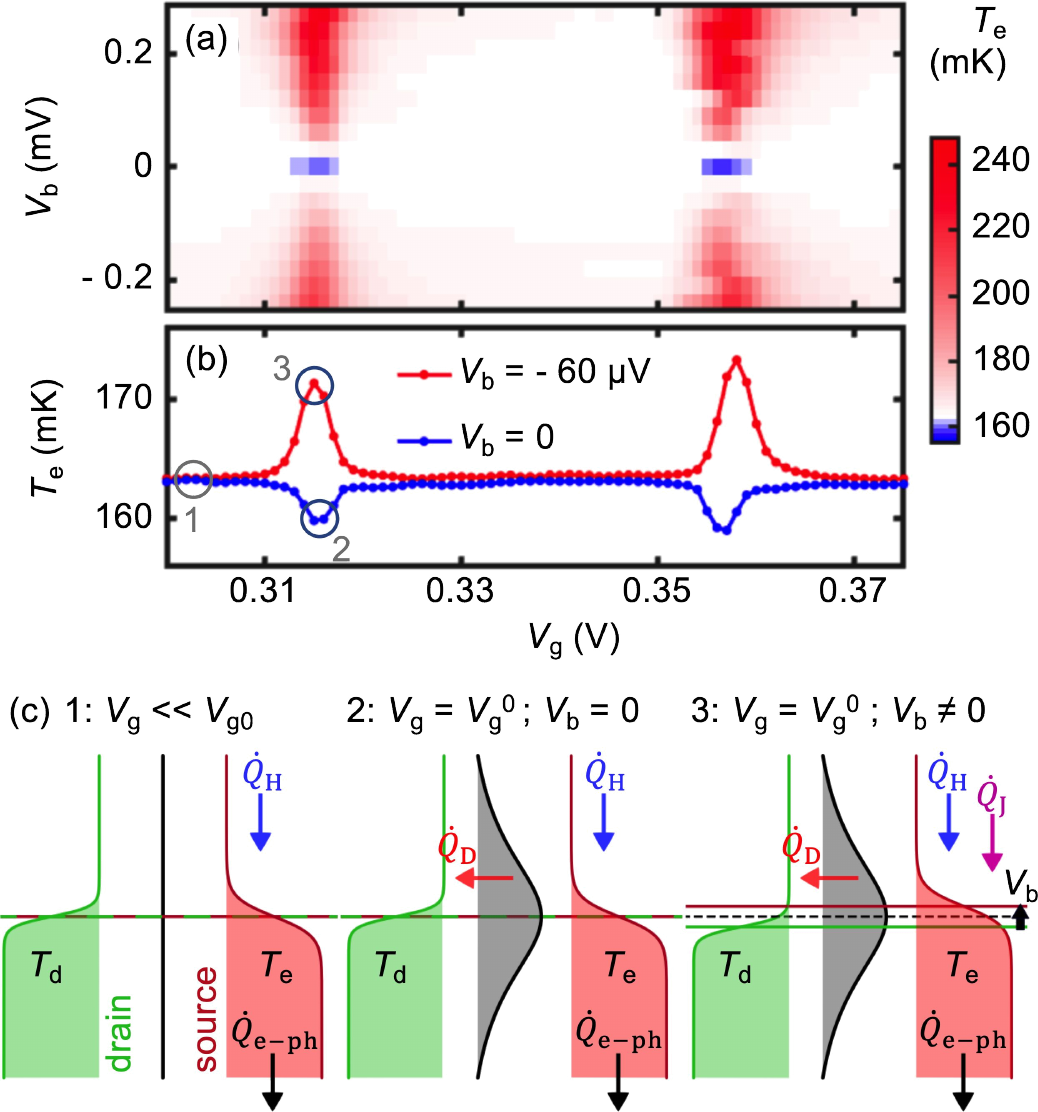}
	\caption{(a) Experimental map of the source electronic temperature in the $\vb-\vg$ plane. (b) Individual gate traces of the source temperature at two different bias values. (c) Schematic energy diagram of the heat flows in/out the source in various conditions as indicated by labels in (b): (1) away from charge degeneracy and at zero bias  (left), (2) at a charge degeneracy point $V_{\rm g} = V_{\rm g}^{\rm 0}$ but still at zero bias (middle) or (3) at non-zero bias in a conducting region (right). The gray profile depicts the quantum level spectral density. The ratio between the level broadening $\hbar \Gamma$, the bias $V_{\rm b}$ and the thermal energy $k_{\rm B}T$ are in correspondence with panel (b) conditions. The arrows indicate the applied heating power $\qdot_{\rm H}$, the Joule power $\qdot_{\rm J}$, the electron-phonon coupling power $\qdot_{\rm e-ph}$ and the power flow through the QD $\qdot_{\rm D}$.}
	\label{fig:figure3}
\end{figure}

In the experiment, we heat up the source by applying a constant heating power $\qdotH=$ 6 fW to the heater junction. The drain is biased at a potential $V_{\rm b}$, the source side being grounded via one of the SNS thermometer contacts. \mbox{Figure \ref{fig:figure3}(a)} shows a map of the source electronic temperature as a function of $V_{\rm b}$ and $V_{\rm g}$. Its resemblance to the charge conductance map of \mbox{Fig. \ref{fig:figure1}(d)} is striking. The source temperature $\te$ increases rapidly with increasing charge current due to the related Joule power. Right at the charge degeneracy point, the source temperature is lower than in the rest of the map. The higher resolution temperature map of \mbox{Fig. \ref{fig:figure4}(a)} shows a clear cooling region of ellipsoidal shape, with slightly canted axes. This cooling effect is the result of the enhanced heat transfer between the hot source and cold drain.

Figure \ref{fig:figure3}(c) shows energy diagrams for three different cases indicated by circles in the temperature $\te(V_{\rm g})$ profiles at two different bias of \mbox{Fig. \ref{fig:figure3}(b)}. At zero bias and far away from charge degeneracy (case 1), there is neither Joule power nor heat flow through the $\qd$. The source is overheated up to $\te=163.5$ mK due to the balance between the applied power $\qdotH$ and the main thermal leakage channel, namely the electron-phonon coupling $\qdot_{\rm e-ph}$. Still at zero bias, but near a charge degeneracy point (case 2), there is a heat flow $\qdot_{\rm D}$ through the $\qd$, but still no charge flow. This shows up (blue curve in \mbox{Fig. \ref{fig:figure3}(b)}) as a temperature $\te$ drop by several mK at the charge degeneracy point. The gate-controlled QD junction thus acts as a heat valve. At higher bias (case 3), this cooling contribution is overcome by the Joule heat $\qdot_{\rm J}$. A temperature maximum is thus observed at values of the gate potential close to the charge degeneracy point (red curve in \mbox{Fig. \ref{fig:figure3}(b)}).

\begin{figure}[!t]
	\includegraphics[width=0.99\columnwidth]{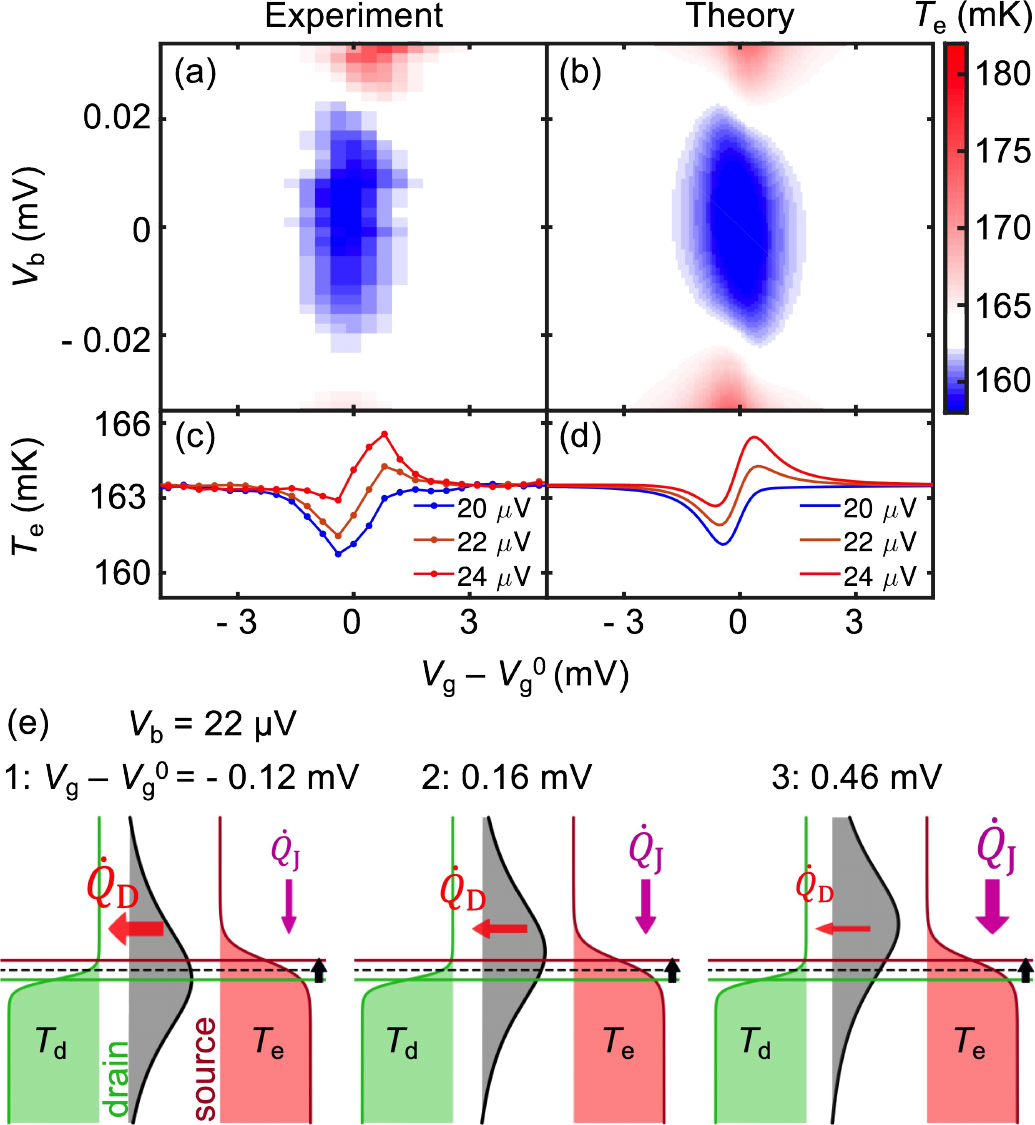}
	\caption{(a) A highly-resolved map of the source electronic temperature at the same experimental condition as in \mbox{Fig. \ref{fig:figure3}} and around a charge degeneracy point defined by $V_{\rm g} = V_{\rm g}^{\rm 0}$. (b) Calculated temperature map obtained with the inbedding technique with $\Gamma$ = 0.25$\mev$, $\Gamma_{\rm L}/\Gamma_{\rm R} =$ 3/17 and $T_{\rm d}$ = 85 mK. (c) Experimental and (d) theoretical variation of the temperature in the region where crossing from cooling to heating is observed; each curve refers to a given applied bias $V_{\rm b}$: (blue) $20 \muv$, (orange) $22 \muv$, (red) $24 \muv$. (e) Schematics describing the crossover between the heat flow $\qdot_{\rm D}$ and the Joule heat $\qdot_{\rm J}$ as a function of the gate at a fixed bias, resulting in temperature decrease at $V_{\rm g} - V_{\rm g}^{\rm  0}$ = - 0.12 mV (case 1, left) or increase at 0.46 mV (3, right). At 0.16 mV (2, middle), the two flows are equilibrated. The electron-phonon heat $\qdot_{\rm e-ph}$ as well as the injected heat $\qdot_{\rm H}$ are omitted for clarity. The widths of the arrows indicate their relative strengths.}
	\label{fig:figure4}
\end{figure}

The mere observation of cooling at the charge degeneracy point is in clear contradiction with the theoretical prediction in the weak coupling, sequential tunneling regime. Indeed the present experiment deals with a strong tunnel coupling between the QD and the leads, with a ratio $\hbar \Gamma/k_{\rm B}T_{\rm e}$$ \approx20$,
rendering the weak coupling picture inapplicable. 

We now go beyond the sequential tunneling approximation. Thanks to the extremely high charging energy, in the vicinity of a charge degeneracy point, the device can be described as a non-interacting single energy level. We are interested in exploring the properties of the leads at stationarity and in particular their electronic temperature; in the NEGFs framework this is possible via the so-called inbedding technique~\cite{stefleebooka,Talarico-PRB20,suppmat}. It is worth mentioning that it is not based on a full heat balance model accounting for the heat flow via phonons and the superconducting leads. We instead assume that the electron-phonon coupling strength itself does not change appreciably within the temperature range of the map, which is equivalent to assume that the main particle and energy redistribution processes in the lead are dominated by electron-electron interactions. By including in the theory the measured temperature (163.5 mK) of the source when decoupled from the $\qd$, we effectively take into account its thermal coupling to the bath.

The theoretical temperature map around a charge degeneracy point is shown in \mbox{Fig. \ref{fig:figure4}(b)} and reveals a nice agreement with the experimental data in \mbox{Fig. \ref{fig:figure4}(a)}. Here, the temperature of the drain $T_{\rm d}$ is set to 85 mK and the coupling of the QD to the drain is asymmetric with a coupling ratio $\Gamma_{\rm L}/\Gamma_{\rm R} =$ 3/17 between left and right leads and $\Gamma$ = 0.25$\mev$. These best fit values allow us to reproduce semi-quantitatively the temperature profiles of the crossing region, see \mbox{Fig. \ref{fig:figure4}(c,d)}. The width in gate potential of the cooling region is independent of the bath temperature and increases with the coupling $\Gamma$ \cite{suppmat}. Conversely its extension in bias depends weakly on $\Gamma$ and increases with the temperature difference across the junction.

The present case actually has some similarities with the regime of a metallic Single Electron Transistor where cooling at the charge degeneracy point was also found \cite{Kubala-PRL08,Dutta-Nanolett19}. Nevertheless, an asymmetry in gate voltage is clearly observed in the experimental and theoretical temperature map. For a bias voltage $V_{\rm b}$ around 22 $\mu V$, the source temperature can be tuned either below or above the reference temperature of 163.5 mK by acting on the gate voltage, see \mbox{Fig. \ref{fig:figure4}}(c). This behavior is not to be expected in the case of a metallic island where electron-hole symmetry in the density of states makes transport properties symmetric across the charge degeneracy point. Therefore it is an unambiguous signature of the QD discrete energy spectrum. At a given bias, the value of the gate potential determines the position of the broadened energy level in the QD (see the grey profile in \mbox{Fig. \ref{fig:figure4}(e)}) and thus the mean energy of the tunneling electrons. This in turn affects the heat balance in the source and modifies the boundary of the cooling region in the temperature map. The extension in bias of this crossover zone, where one can switch from cooling to heating by adjusting with the gate, depends on both the coupling $\Gamma$ and the temperature difference across the QD \cite{suppmat}. 

This work shows that electronic heat transport through a QD junction can be modulated by a gate potential, making it act as a gate-tunable heat valve. This behavior can have important consequences in the practical thermo-electric efficiency of such a single quantum-dot junction \cite{Harzheim-PRR20}. The Coulomb diamond patterns in the temperature maps reveal the intimate relation between charge conductance on one hand and heat transport and dissipation on the other. Further experiments may allow a quantitative comparison of thermal effects to the charge transport properties, in a wide range of tunnel couplings. The ability of precision electron thermometry at the heart of a QD-based device demonstrated here opens wide perspectives in the field of heat transport and dissipation in quantum electronic devices, paving the way for quantitative tests of the Landauer principle in the quantum information regime \cite{RevuePekola}. 

\begin{acknowledgments}
The authors thank \mbox{B.} Karimi and \mbox{J. P.} Pekola for stimulating discussions. Samples were realized at the Nanofab platform at Institut N\'eel with the help of \mbox{T.} Crozes. We acknowledge support from the Nanosciences Foundation under the auspices of the Foundation UGA and from the European Union under the Marie Sk\l odowska-Curie Grant Agreement 766025. NLG and WNT acknowledge support from the Academy of Finland Center of Excellence program (Project \mbox{no.} 312058) and the Academy of Finland (Project \mbox{no.} 287750). NLG acknowledges funding from the Maupertuis Program for a RSM grant and from the Turku Collegium for Science and Medicine. Numerical simulations were performed at the Finnish CSC facilities under the Project \mbox{no.} 2000962.
\end{acknowledgments}

\clearpage
\widetext
\begin{center}
\textbf{\large Supplemental Material: A Single-Quantum-Dot Heat Valve}
\end{center}
\setcounter{equation}{0}
\setcounter{figure}{0}
\setcounter{table}{0}
\setcounter{page}{1}
\makeatletter
\renewcommand{\theequation}{S\arabic{equation}}
\renewcommand{\thefigure}{S\arabic{figure}}
\renewcommand{\bibnumfmt}[1]{[S#1]}
\renewcommand{\citenumfont}[1]{S#1}
\newcommand{\ep}{\epsilon}
\newcommand{\vep}{\varepsilon}
\newcommand{\hc}{{\rm \;h.\,c.\;}}
\newcommand{\sign}{\mathop{\mathrm{sign}}\nolimits}
\renewcommand{\Im}{\mathop{\mathrm{Im}}\nolimits}
\renewcommand{\Re}{\mathop{\mathrm{Re}}\nolimits}



\title{Supplemental Material: A Single-Quantum-Dot Heat Valve}

\author{B. Dutta}
\affiliation{\mbox{Univ.} Grenoble Alpes, CNRS, Grenoble INP, Institut N\' eel, 25 Avenue des Martyrs, 38042 Grenoble, France}
\author{D. Majidi}
\affiliation{\mbox{Univ.} Grenoble Alpes, CNRS, Grenoble INP, Institut N\' eel, 25 Avenue des Martyrs, 38042 Grenoble, France}
\author{N. W. Talarico}
\affiliation{QTF Centre of Excellence, Turku Centre for Quantum Physics, Department of Physics and Astronomy, University of Turku, 20014 Turku, Finland}
\author{N. Lo Gullo}
\affiliation{QTF Centre of Excellence, Turku Centre for Quantum Physics, Department of Physics and Astronomy, University of Turku, 20014 Turku, Finland}
\author{C. B. Winkelmann}
\affiliation{\mbox{Univ.} Grenoble Alpes, CNRS, Grenoble INP, Institut N\' eel, 25 Avenue des Martyrs, 38042 Grenoble, France}
\author{H. Courtois}
\affiliation{\mbox{Univ.} Grenoble Alpes, CNRS, Grenoble INP, Institut N\' eel, 25 Avenue des Martyrs, 38042 Grenoble, France}


\date{\today}
\maketitle
In this supplemental material part, we discuss about the sample fabrication, charge  transport properties, performance of our $SNS$ thermometer junction as a bolometric detector, and the details of the theoretical approach used to explain our data.



\section{Sample fabrication and charge transport characteristics}

We have used a Si substrate covered with a 300 nm SiO$_2$ oxide, from which gold films thinner than about 10 nm are known to dewet. Yet, and most interestingly, the gold nanoparticles even happen to form on top of the noble metal surfaces (Au, Pt) from which the electromigration constriction is formed. This means that a thin gold film does not wet on gold. This was also observed by the Cornell group \cite{Bolotin2004}. In line with these authors, we believe that even after thorough pumping down to a few $10^{-7}$ mbar (as is usual in evaporation chambers), the noble metal surfaces are still not clear of a contaminant layer (\mbox{e.g.} water), which leads to dewetting of the nm-thin gold top layer.

The samples in \mbox{Ref.} \cite{Dutta-Nanolett19} and in the present manuscript share the same geometry and fabrication technique but they are different samples that were fabricated in two different runs. The formation of a nano-gap by electromigration depends sensitively of the structural details (precise width and thickness) of the constriction, therefore the size of the gap (between leads and dot) and its structural details vary from one electromigration to other. As a result, the strength of tunnel coupling also varies. Here, the tunnel coupling in \mbox{Ref.} \cite{Dutta-Nanolett19} device was about 10 times larger than in the present device. In both cases Pt was used as the electro-migration material, as it suppresses the superconducting proximity effect extremely efficiently, much more than Au.

The charging energy of the dot depends on the actual size of the Au nano-island and the total effective capacitance with its environment, that is determined by both the precise nature of the evaporated Au droplets and the detailed structure of the nano-gap created by electromigration. Therefore, it is expected to have a different charging energy for two otherwise similar samples. Metallic quantum dots have been investigated in the 90s. In \mbox{Ref. \cite{Ralph-PRL95}}, an energy level separation of 0.7 meV and a charging energy of 6 meV were deduced from the measured energy spectra in 10 nm Al nanoparticles. The nanoparticles used in our work can be seen from SEM images to be about 5 nm in size and we expect an energy level separation of the order of a few meV. In a previous work \cite{vanZanten-PRB16} from our group, the weak coupling to leads enabled us to observe sharp resonances in the differential conductance map corresponding to an energy level separation of about 5 meV. 

Figure 1 of the main text displays the differential conductance map as a function of both the bias and gate voltages $V_{\rm b}$ and $V_{\rm g}$ respectively. A Coulomb diamond pattern is clearly, but partly, visible. The charging energy of 4 meV is estimated from extrapolating the bias at the top of a diamond, which is actually twice the charging energy \cite{Thijssen-PSSB08}. We did not measure this device above bias voltages of about $\pm$ 4 mV, as, due to the rather large tunnel couplings, this voltage leads to currents of about 6 nA, beyond which we observe instabilities related to inelastic processes. At higher bias, there is a risk of burning the device. Therefore, a full spectroscopic characterization revealing several successive levels was not possible.

The detailed shape of the Coulomb diamond pattern can be used to determine the different capacitive couplings of the QD to its environment. In particular the sum of the inverse of the diamond (positive and negative) slopes is equal to the ratio of the total capacitance to its leads over the capacitance to the gate $\alpha^{-1}$ \cite{Bonet-2002}. The so-called coupling parameter $\alpha$ translates the effect of the gate voltage in terms of shift in chemical potential of the QD. Here we obtain $\alpha\approx$ 0.157.

\begin{figure}[t]
	\includegraphics[width=0.6\columnwidth]{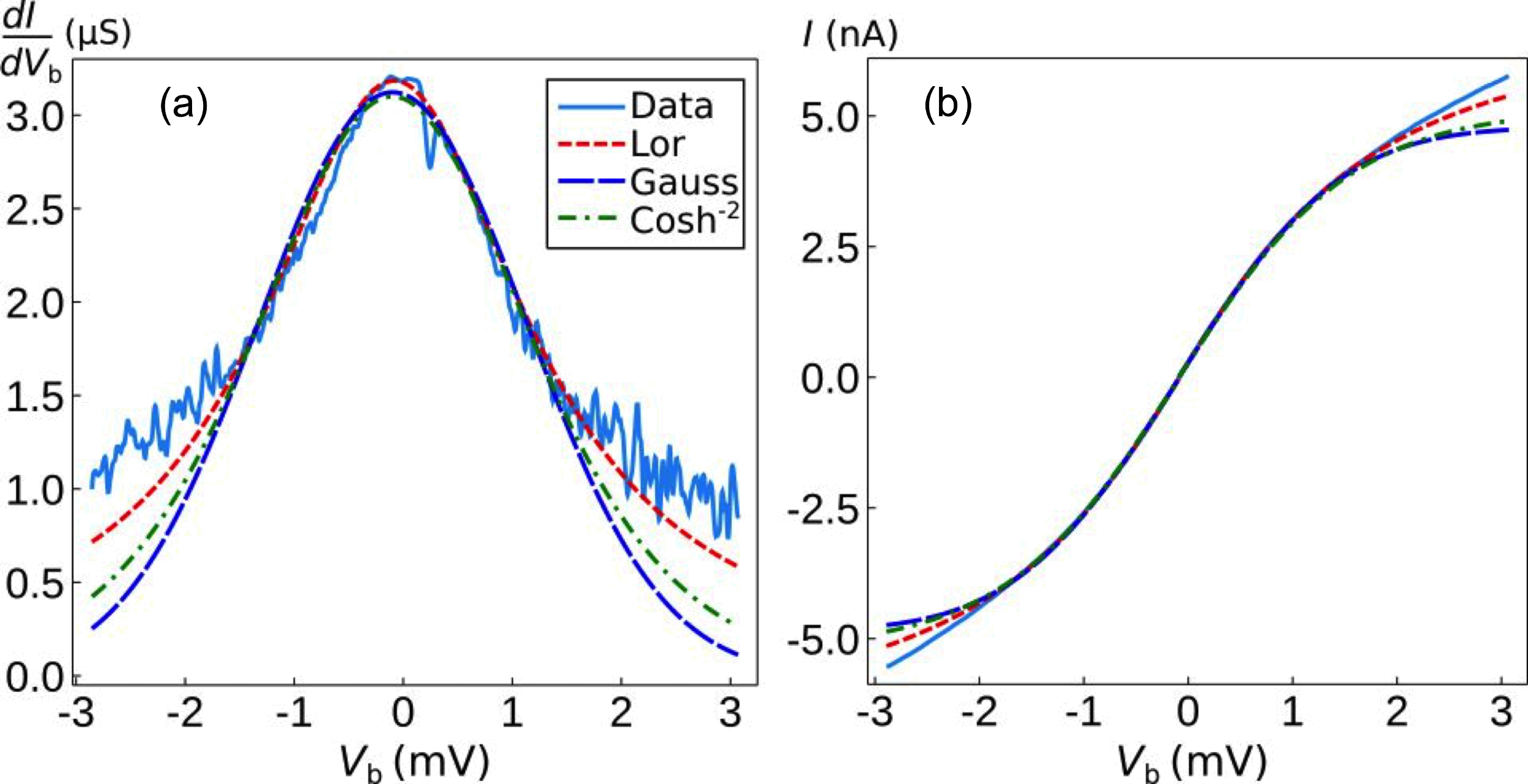}
	\caption{(Color online) Measured (blue lines) differential conductance $dI/dV_{\rm b}(V_{\rm b})$ at the charge degeneracy point $V_{\rm g}=-1.581 \mv$ (panel a) and current $I(V_{\rm b})$ computed from it (panel b) compared with that theoretically predicted by \mbox{Eq.~\ref{eq:sigma}} at fixed gate voltage and as a function of the applied bias voltage (red dotted lines). Alternative fits with a Gaussian (blue dotted line) or a $cosh^{-2}$ (green dotted line) profile for the differential conductance are also shown.}
	\label{fig:sV-IV}	
\end{figure}

We now discuss the nature of charge transport through the QD. This can be done by means of the differential conductance data shown in \mbox{Fig. 1} (d) of the main text. In Fig.~\ref{fig:sV-IV} we show the measured differential conductance $dI/dV_{\rm b}(V_{\rm  g},V_{\rm b})$ (left) and the corresponding current (right) obtained by integrating it, both at the charge degeneracy point $V_{\rm g}=-1.581 \mv$.

We first compare the measured differential conductance to the theoretical expression for the charge transport through a single quantum level:
\begin{equation}
\frac{dI}{dV_{\rm b}}(V_{\rm g},V_{\rm b})=\int \frac {d\omega}{2\pi} \frac{\Gamma_{\rm s} \Gamma_{\rm d}}{[\omega-e(V_{\rm g}-V_{\rm g}^{\rm 0})]^2+(\Gamma_{\rm T}/2)^2} \frac{dn(\omega)}{dV_{\rm b}}
\label{eq:sigma}
\end{equation}
where $n(\omega)$ is the Fermi-Dirac distribution. The total broadening is given by $\Gamma_{\rm T}=\Gamma+\Gamma_{\rm ext}$ where $\Gamma=\Gamma_{\rm s}+\Gamma_{\rm d}$ is induced by the coupling to the leads whereas $\Gamma_{\rm ext}$ is added to account for a possible extra broadening mechanism such as fluctuations of the applied gate voltage. For now we assume $\Gamma_{\rm ext}=0$. Around the resonance $V_{\rm g}=V_{\rm g}^{\rm 0}$, the differential conductance has a Lorentzian shape with a width given by $2\Gamma_{\rm T}$ and a maximum determined by the ratio $\Gamma_{\rm s}/\Gamma_{\rm d}$. \mbox{Figure \ref{fig:sV-IV}(a)} shows, in parallel with the experimental data, the best fit with the two free fit parameters $\Gamma_{\rm T}=\Gamma\approx 1.5 \mev$ and $\Gamma_{\rm s}/\Gamma_{\rm d}=0.16$. The agreement is excellent. Here the temperature of both leads set to $T_{\rm b}=100 \mk$ does not contribute significantly to the obtained lineshape. In addition, \mbox{Fig. \ref{fig:sV-IV}} includes alternative fits with a Gaussian or a $cosh^{-2}$ profile for the differential conductance. These fits are much less in accordance with the data, which strengthens the above conclusion.

The above analysis demonstrates that electron transport occurs through a single-level quantum dot. The observed Lorentzian line-shape is a strong indication that the main mechanism for the broadening of the spectral function of the dot is due to the coupling with the leads. Any other mechanism, if present, contributes only marginally.\\

\section{Performance of a SNS junction as a bolometer}

In order to measure the heat flow through a QD junction, which is described in the main text, one needs to be able to access a very small change in electronic temperature. Moreover, one needs an operating temperature of down to 100 mK or below, where the QD heat flow dominates over the other paths of heat relaxation such as electron-phonon coupling. These two requirements lead us to consider SNS proximity junction as a thermometer that can fulfill both of these requirements. We have optimized the sensitivity of the SNS thermometer with several repetitions of the junctions parameters such as the length and thickness of the normal metal. In this way, we reduced the Thouless energy ($\ethls$) of the SNS junction, which basically determines the lowest saturation temperature of the thermometer \cite{Dubos-PRB01b}.

\begin{figure}[t]
	\includegraphics[width=0.6\columnwidth]{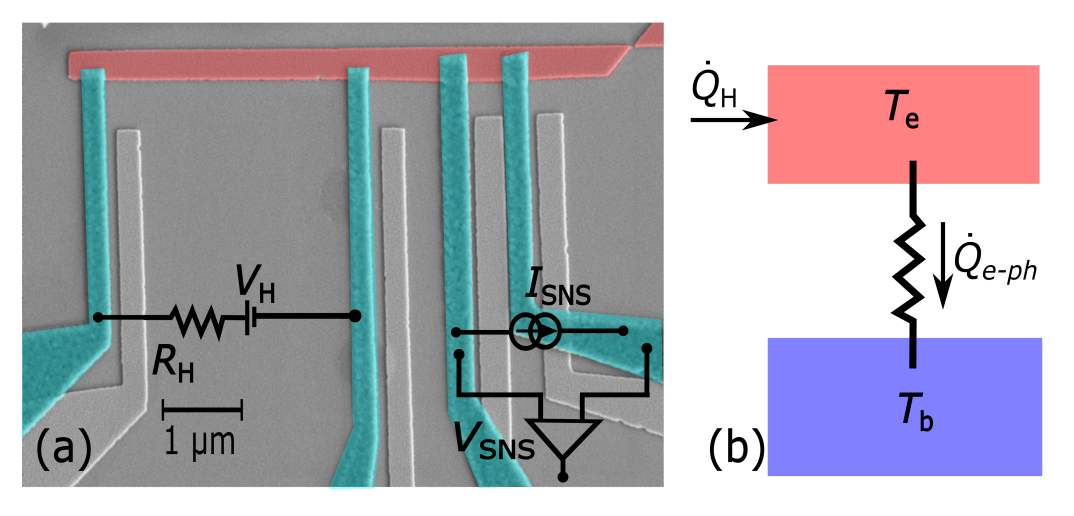}
	\caption{(a) Colored scanning electron micrograph of one of our sample. The source side is separated from the drain by electromigration and there is no quantum dot placed on the junction. Therefore it can be considered as a small metallic island with a heater junction (two Al leads on the left) and a thermometer junction (two Al leads on the right). (b) Equivalent thermal model of this device, showing how the injected heat from the heater is equilibrated via electron-phonon coupling.} \label{fig:SNSBolometer}
\end{figure}

Here we describe a test experiment, where we determine the sensitivity of our optimized SNS thermometer and test its operation as a bolometric detector. The SEM image of the device under test is shown in \mbox{Fig. \ref{fig:SNSBolometer}(a)}, where the normal metal Au is shown in red color and the superconducting Al leads in light-blue. The basic structure of the device is similar to that of the QD device described in the main text, with  the only difference that there is no QD is placed in between source and drain after the electromigration. Therefore, the device can be essentially considered as a $\sim$ 5 $\mu$m long and $\sim$ 100 nm wide rectangular normal metallic island. Like the samples discussed in the main text it has a very long SNS junction to inject Joule heat into it and a short SNS junction to measure the electronic temperature.

\begin{figure}[t]
	\includegraphics[width=0.6\columnwidth]{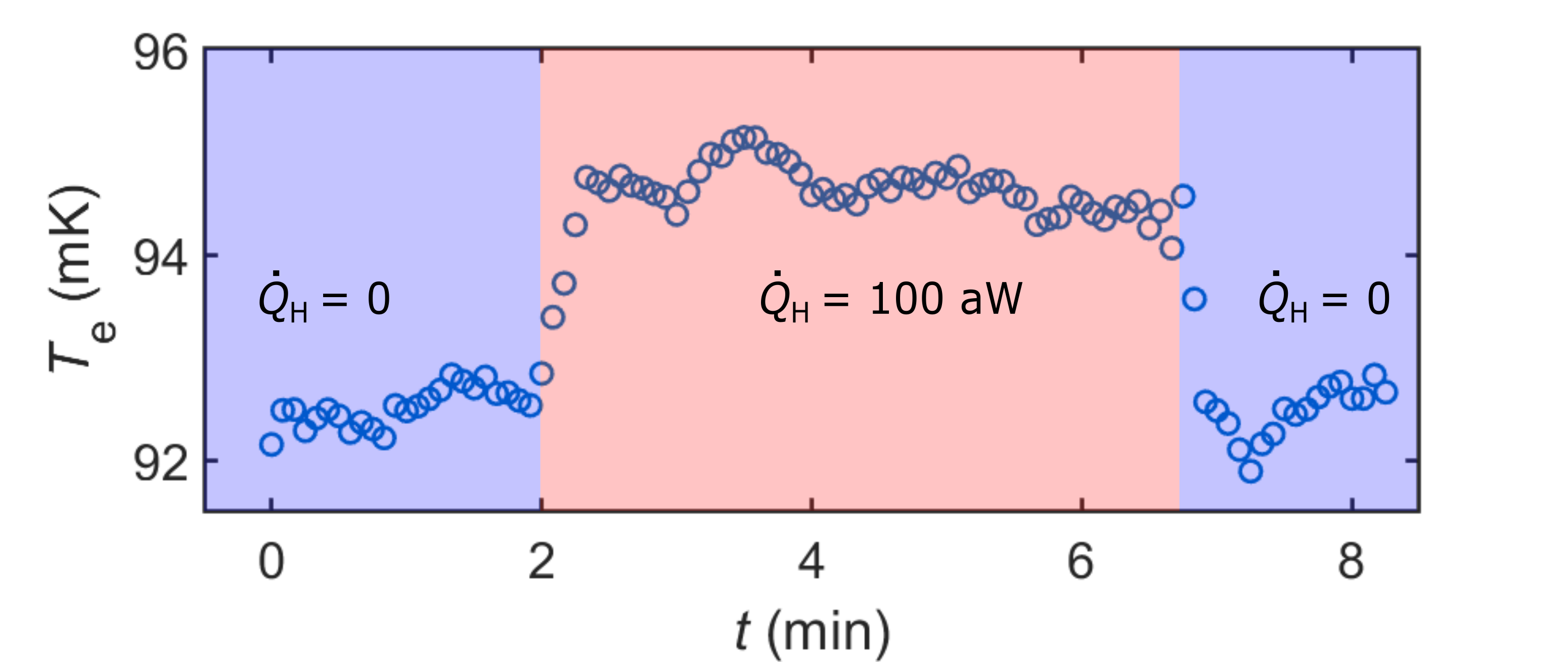}
	\caption{A real-time measurement of the electronic temperature of the source island. Each point is a Gaussian maximum of the histogram of 500 measurements of stochastic switching current, taken in 1 sec. One can easily notice a change of the electron temperature by a few mK compared to the background temperature of 93 mK, whenever we turn on (off) the heater, set to an input heating power of 100 aW.} \label{fig:Te_timetrace}
\end{figure}

We heat-up the island by applying a constant adjustable \mbox{d.c.} current through the heater junction, using a 1.3 V isolated \mbox{d.c.} battery. The SNS thermometer is calibrated against the well known bath temperature, by measuring the histograms of its stochastic switching current, as described in the main text. Here we present an experiment, where the bath temperature is at $T_{\rm b}$ = 90 mK and the heater junction is current-biased through a 200 M$\Omega$ biasing resistor, which leads to a heating power $\dot{Q}_{\rm H}$ = 100 aW. We continuously monitor the electronic temperature of the island by measuring a histogram of 500 switching currents in about 1 sec. The real time temperature trace of the island is shown in \mbox{Fig. \ref{fig:Te_timetrace}}. One can easily identify the change of the electronic temperature by a few mK \mbox{w.r.t.} the background temperature of about 92 mK, whenever the heater is turned on (off). Therefore the thermometer clearly detects an input heat as low as 100 aW, thus performing as a bolometric detector of very small heating power. The noise equivalent power is about 100 aW/$\sqrt{\rm Hz}$.

This observation of the island's electronic temperature can be determined by a heat-balance equation, as shown by a heat-balance model in the \mbox{Fig. \ref{fig:SNSBolometer}(b)}: 
\be
\dot{Q}_{\rm H} - \Sigma \mathcal{V} (T_{\rm e}^{5}-T_{\rm b}^{5}) = 0,
\label{eq:heat-balance}
\ee
where $\Sigma$ is the material dependent constant, $\mathcal{V}$ is the volume of the island, $T_{\rm e}$ and $T_{\rm b}$ are the electron and the bath (phonon) temperatures respectively. Any parasitic heat source (sink) such as heat losses through the superconducting leads due to imperfect thermal insulation \cite{Joonas-PRL2010} or parasitic heating by the electromagnetic environment are taken into account within the injected heating power $\dot{Q}_{\rm H}$.

The electronic temperature of the island can be extracted by solving the above heat-balance equation (\ref{eq:heat-balance}). If we use an injected heating power $\dot{Q}_{\rm H}$ = 100 aW, the material constant for Au $\Sigma = 2.4 \times 10^{9}$ Wm$^{-3}$K$^{-5}$ \cite{Giazotto-RMP06}, the volume of the island $\mathcal{V} = 2 \times 10^{-20}$ m$^{3}$ and the bath temperature $T_{\rm b} = $ 80 mK, we get an increase of the electronic temperature $\Delta T_{\rm e} \sim$ 3 mK, which is consistent with the measured value. This justifies the analysis of the heat relaxation mechanism in the island as discussed above. 

In the experiment described in the main text, we use a saw-tooth shaped \mbox{a.c.} current bias of the junction to measure 3000 switching events in about 10 sec. The critical current of the junction is determined as the gaussian maximum. The thermometer sensitivity is found to be 1.5 $\mu$A/K at 80 mK, with a noise level of about 200 $\mu$K/$\sqrt{\rm Hz}$.

\section{Inbedding technique}

Here, we give a brief overview of the inbedding technique used to compute the non-equilibrium steady-state of the leads and their electronic temperatures under the hot-electron assumption.
The Hamiltonian of the total system (quantum dot plus source and drain reservoirs) is given by:
\begin{eqnarray}
\hat H&=\hat H_\qd+\underset{\alpha=s,d}{\sum}\hat H_{\alpha} +\underset{\rm \alpha=S,D}{\sum}\hat V_{\qd}^{(\alpha)} \\
\hat H_\qd&=v_{\rm g} \hat d^\dag \hat d\;,\hspace{0.5cm}\hat{H}_\alpha=\underset{\rm k_\alpha}{\sum}\epsilon_{\rm k_\alpha}\hat c^\dag_{\rm k_\alpha} \hat c_{\rm k_\alpha} \\
\hat V_{\qd}^{(\alpha)}&=g_\alpha \underset{\rm k_\alpha}{\sum}\bigg(\hat c^\dag_{\rm k_\alpha} \hat d+\hat d^\dag \hat c_{\rm k_\alpha}\bigg).
\end{eqnarray}
where $v_g=\alpha(V_g-V_g^0)$ is the gate voltage measured from the considered resonance and accounting for the coupling parameter. $V_{\qd}^{(\alpha)} $ the coupling Hamiltonian between the quantum dot, $\qd$, and the $\alpha=s,d$ leads. 
Using the non-equilibrium Green's function approach \cite{stefleebook}, and assuming that the whole system reaches a (possibly non-equilibrium) stationary state, the state of the $\qd$ is completely characterized by the retarded and lesser single-particle Green's functions:
\begin{eqnarray}
\label{eq:dysons}
G^R_{\qd}(\omega)&=\left(\mathds{1}-g^R_{\qd}(\omega)\Sigma^R_{emb}(\omega)\right)^{-1}g^R_{\qd}(\omega)\\ G^<_{\qd}(\omega)&=G^R_{\qd}(\omega)\Sigma^<_{emb}(\omega)G^A_{\qd}(\omega),
\end{eqnarray}
here $G^A_{\qd}(\omega)=[G^R_{\qd}(\omega)]^\dag$ is the advanced component of the Green's functions. $g^R$ is the Fourier transform of the retarded Green's function of the isolated (non-coupled to the leads) quantum dot $g_\qd(t,t')=i\theta(t-t')\langle[\hat d(t),\hat d^\dag(t')]\rangle$, with $\hat d(t)=e^{i\hat{H}_\qd t}\hat d e^{-i\hat{H}_\qd t}$. The embedding self-energy is defined  as $\Sigma^K_{emb}(\omega)=\underset{\alpha}{\sum}|g_\alpha|^2 g_\alpha^K(\omega)$, with $K=R,<$ and $g_\alpha^K(\omega)$ the Fourier transform of isolated leads' Green's functions.
We work in the wide band limit approximation (WBLA) and therefore we have:  $\Sigma^R_{emb}(\omega)=-i\underset{\alpha}{\sum}|g_\alpha|^2 /2=-i\Gamma/2$.
The first equation in Eqs.~\ref{eq:dysons} is the Dyson equation for the retarded component of the single-particle Green's function, from which the spectral function of the $\qd$ can be computed as $A_{\qd}(\omega)=-\pi^{-1} \text{Im} G^R(\omega)$.

\begin{figure}[t!]
	\includegraphics[width=0.75\linewidth]{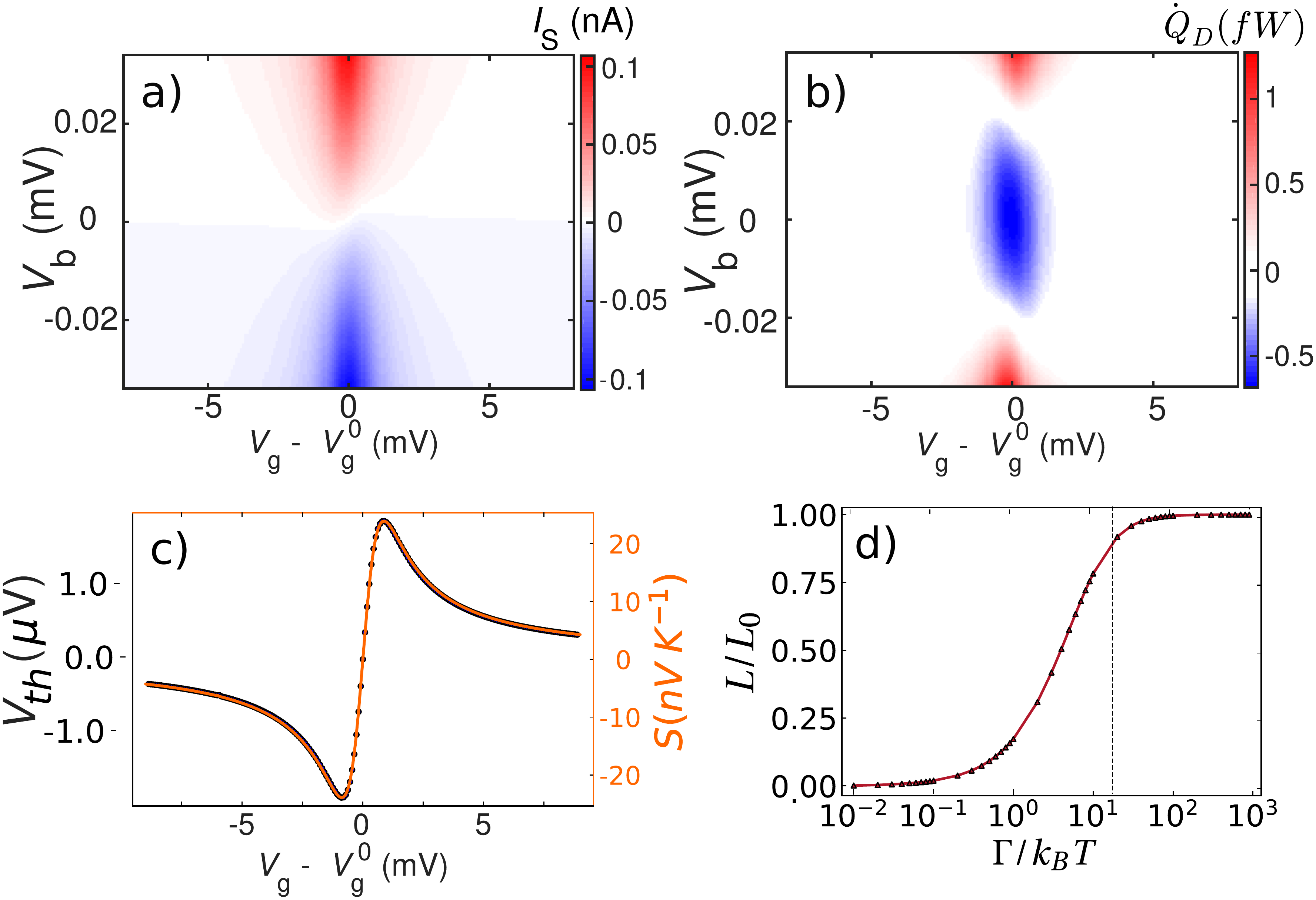}
	\caption{(Color online) Maps of the a) particle and b) heat current for the source lead. c) Thermovoltage $V_{\rm th}$ and corresponding thermopower $S=V_{\rm th}/(\Delta T)$ as a function of the gate voltage $V_{\rm g}$. The parameters of the system are the same as the one considered in the main text: coupling is $\Gamma=0.25 \mev$, and temperature of the drain at closed bias is $T_{\rm d}=85$ mK. (d) Lorentz ratio $L/L_{\rm 0}$ as a function of $\Gamma/k_{\rm B} T$ for a single level quantum dot. The dashed lines shows the ratio for the dot studied in the main text $\Gamma/k_{\rm B} T\approx 20$ and the dot is at the charge degeneracy point $V_{\rm g}=V_{\rm g}^{\rm 0}$.}
	\label{fig:maps}
\end{figure}

The embedding self-energy approach accounts for the effect that the leads have on the physical properties of the system, but once the solution to Eqs.~\ref{eq:dysons} are obtained, it is possible to find the Green's functions of the leads and explore the influence of the $\qd$ on the physical features of the reservoirs. By introducing the inbedding self-energy $\Sigma^K_{\rm inb,\alpha}(\omega)=|g_\alpha|^2 G_{\qd}^K(\omega)$ one has the following relations:
\begin{eqnarray}
\label{eq:dysonsleads}
&G^R_{\alpha}(\omega)=g^R_{\alpha}(\omega)+g^R_{\alpha}(\omega)\Sigma^R_{\rm inb,\alpha}(\omega)g^R_{\alpha}(\omega)\\ 
&G^<_{\alpha}(\omega)=g^<_{\alpha}(\omega)+g^r_{\alpha}(\omega)\Sigma^<_{\rm inb,\alpha}(\omega)g^A_{\alpha}(\omega)\\
&+g^<_{\alpha}(\omega)\Sigma^A_{\rm inb,\alpha}(\omega)g^A_{\alpha}(\omega)+g^r_{\alpha}(\omega)\Sigma^R_{\rm inb,\alpha}(\omega)g^<_{\alpha}(\omega).\nonumber
\end{eqnarray}

The lesser Green's functions of the leads $G_\alpha^<(\omega)$ are different from the isolated ones $g_\alpha^<(\omega)=i f_\alpha(\omega)$, with $f_\alpha(\omega)$ the Fermi-Dirac distributions characterized by the initial chemical-potential and temperature $\mu_\alpha$ and $T_\alpha$.
From the knowledge of $G_\alpha^<(\omega)$ we can compute both the average particle number and energy of the leads:
$N_\alpha=i\int \frac{d\omega}{2\pi} G^<_{\alpha}(\omega)$ and $E_\alpha=-i\int \frac{d\omega}{2\pi} \omega\;G^<_{\alpha}(\omega)$.

We now resort to the hot electron assumption that accounts for a neat separation of the energy relaxation time scales between system and leads. Basically, it assumes that the electron-electron interactions into the leads makes their equilibration time much faster than any other dynamical processes in the whole system. Therefore, we can assume that the leads at the steady state  are described by a new Fermi-Dirac distribution $f_{\alpha}'(\omega)$ with a different set of chemical potential and temperature $\mu_\alpha'$ and $T_\alpha'$. The two parameters are the solution of the set of non-linear equations: 

\begin{eqnarray}
\label{eq:nnlinearsys}
&\int \frac{d\omega}{2\pi} f_{\alpha}'(\omega)=N_\alpha \\
&\int \frac{d\omega}{2\pi} \omega\;f_{\alpha}'(\omega)=E_\alpha.
\end{eqnarray}

\section{Particle and energy currents}

Within the non-equilibrium Green's function formalism, particle, energy and heat currents of the source lead are given by the following expressions:
\begin{eqnarray}
\label{eq:currents}
\label{eq:pcur}
&I_{\rm s}= \int\frac{d\omega}{2\pi}\;\mathcal{T}_{sd}(\omega)(f_{\rm s}(\omega)-f_{\rm d}(\omega))\\ 
\label{eq:ecur}
&J_{\rm s}= \int\frac{d\omega}{2\pi}\;\omega \;\mathcal{T}_{sd}(\omega)(f_{\rm s}(\omega)-f_{\rm d}(\omega))\\
\label{eq:hcur}
&\dot Q_{\rm s}=J_{\rm s}-\mu_{\rm s} I_{\rm s},
\end{eqnarray}
where we used the definition of the transmission coefficient in the WBLA $\mathcal{T}_{\rm sd}(\omega)=\Gamma_{\rm s}\Gamma_{\rm d}\textrm{Tr}[G^{\rm R}(\omega)G^{\rm A}(\omega)]$ which is valid beyond the single level approximation for the dot.
For the set of parameters considered in the main text, the heat and particle currents for the source are shown in \mbox{Fig. \ref{fig:maps}}. It is interesting to notice the resemblance between the heat current map and the temperature map shown in the main text. It confirms that in the regime we explored the temperature changes in the source lead correspond indeed to a heat current to/from the source (panel (a) in \mbox{Fig. \ref{fig:maps}}). At large bias the source heats up; the system behaves as a heater, namely the energy of the bias is transformed into internal energy. At low bias we observe instead heat flow from the hot to the cold lead; the system behaves as a valve, meaning that it enables the natural flow of heat from the hot to the cold lead. It is also interesting to observe that there is a whole region in which heat and particle currents have opposite signs.

It is worth to mention that this effect is not due to the onset of a thermovoltage which would make particles flow against the applied bias voltage without necessarily causing an inversion of heat current. The thermovoltage, although present, is very small compared to the extension in bias voltage where the mismatch in sign is observed. This is shown in \mbox{Fig. \ref{fig:maps}} panel (c) where we plot the thermovoltage $V_{\rm th}(V_{\rm g})$, together with the corresponding thermopower, defined as the bias voltage at which the particle current vanishes.

The observed significant thermal conductance is a signature of the strong coupling of the QD to the leads. We computed the Lorentz ratio $L=\kappa/(T\sigma)$ where $\kappa=\partial\dot Q_{\rm s}/{\partial \Delta T}|_{I_{\rm s}=0}$ and $\sigma=\partial I_{\rm s}/{\partial V_{\rm b}}|_{\Delta T=0}$ are the thermal and electrical conductivities respectively. In panel (d) of Fig. \ref{fig:maps} we plot the Lorentz number $L/L_{\rm 0}$ with $L_{\rm 0}=(\pi^2/3)(k_{\rm B}/e)^2$ at the charge degeneracy point $V_{\rm g}=0$. The dashed line corresponds to the ratio $\Gamma/k_{\rm B} T\approx20$ considered in the system presented in the main text. It is clear that the deviation from the WF law is small because of the strong coupling to the leads.

\section{The cooling and transition regions}

\begin{figure}[!t]
	\includegraphics[width=0.6\columnwidth]{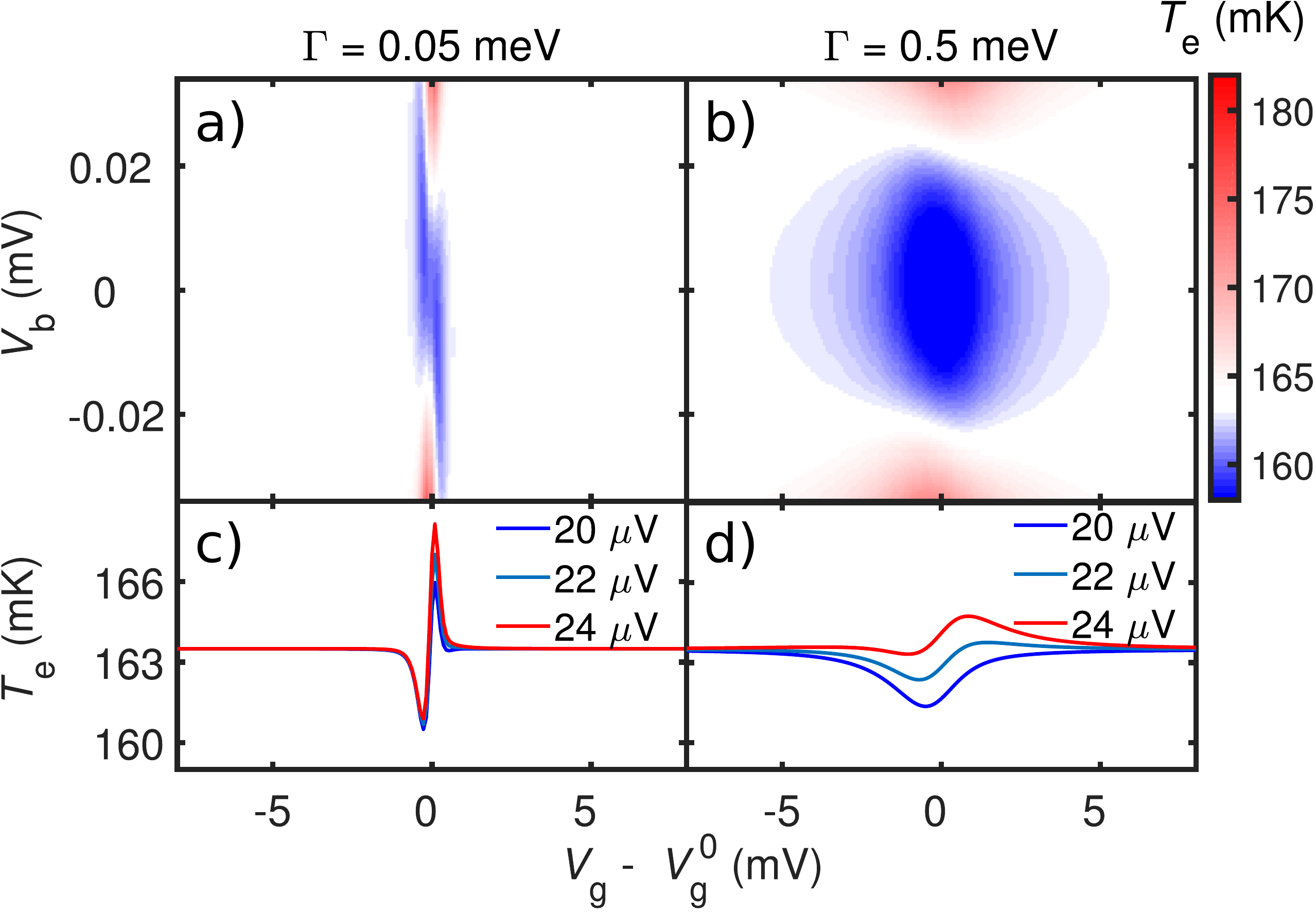}
	\caption{Zoomed temperature map: Calculated temperature map of the source lead obtained with the inbedding technique with (a) $\Gamma$ = 0.05 $\mev$ and (b) $\Gamma$ = 0.5 $\mev$. (c) and (d) variation of the temperature in the region where crossing from cooling to heating is observed; each curve refers to a given applied bias $V_{\rm b}$: (blue) $20 \muv$, (cyan) $22 \muv$, (red) $24 \muv$.}
	\label{fig:cooling}
\end{figure}

\mbox{Figure \ref{fig:cooling}} shows the map of the calculated electronic temperature for couplings (panels (a) and (c)) $\Gamma=0.05 \mev$ and (panels (b) and (d)) $\Gamma=0.5 \mev$ and for the same drain temperature at closed gate voltage $T_{\rm d}=85$ mK similarly to \mbox{Fig. 3} of the main text. A change in the coupling changes the extension of the cooling region in the gate voltage but it does not affect dramatically the extension of the cooling region in the bias voltage nor the position of the transition from cooling to heating. Compared to the discussion in the main text and in Figure \ref{fig:cooling}, we consider a symmetric coupling of the dot to the leads, which increases the amplitude of the cooling.

\begin{figure*}[t]
	\includegraphics[width=0.9\columnwidth]{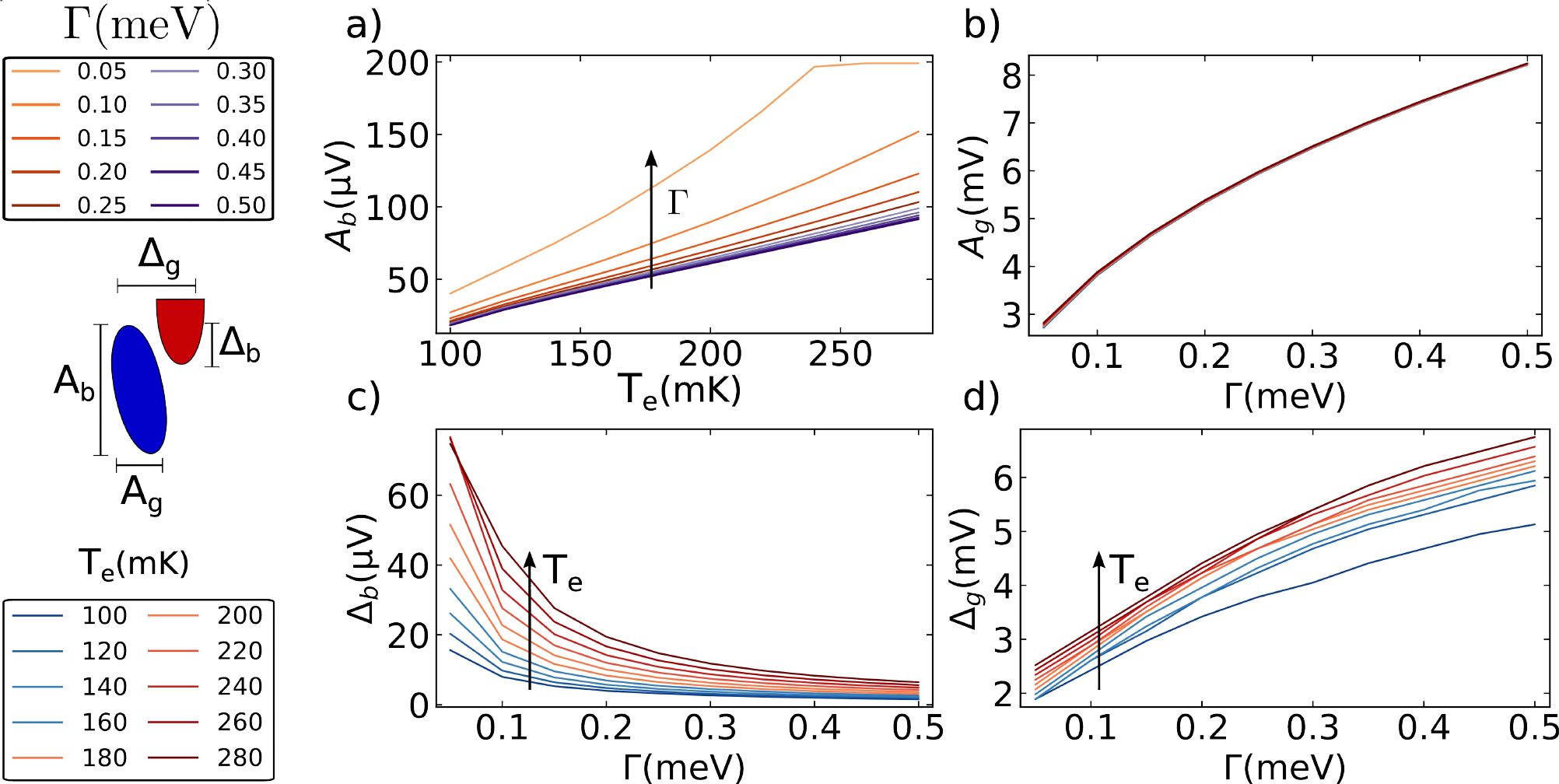}
	\caption{Cooling region characterization: (a) and (b) show the width in gate potential and the extension in bias voltage of the cooling region as a function of the coupling $\Gamma$ for different values of the temperature of the source lead at closed gate. (c) and (d) show the width in gate potential and the extension in bias voltage of the region in which cooling and heating can be observed at fixed bias by tuning the gate voltage as a function of the coupling $\Gamma$ for different values of the temperature of the source at closed gate.}
	\label{fig:coolregion}
\end{figure*}

To give a more quantitative analysis we computed the extension of the cooling region in the bias voltage $A_{\rm b}$ at $V_{\rm b}=0$ and its width $A_{\rm g}$ in gate voltage for different couplings and temperatures of the drain at closed gate. The results are plotted in \mbox{Fig. \ref{fig:coolregion}} in panels (a) and (b) where it can be appreciated that the coupling constant does not change significantly the extension in bias which instead strongly depends upon the difference in the equilibrium temperatures between the drain and source. Indeed as the temperature of the drain increases towards the temperature of the source the extension in $V_{\rm b}$ shrinks. Nevertheless the extension in the gate voltage is only determined by the coupling and does not present any significant dependence upon the closed gate temperature of the drain.

The transition region, namely the region where at fixed bias it is possible to obtain both heating and cooling by changing the gate potential, has a strong dependence on both the coupling and the temperature difference at closed gate. This is shown in \mbox{Fig. \ref{fig:coolregion}} panels (c) and (d) where we plot the width $\Delta_{\rm g}$ and extension $\Delta_{\rm b}$ of the transition region; they have been determined by finding the curve $V_{\rm b}(V_{\rm g})$ such that $T_{\rm e}=$163.5 mK and then taking $\Delta_{\rm b}=\text{Max} (V_{\rm b}(V_{\rm g}))-\text{min} (V_{\rm b}(V_{\rm g}))$ whereas $\Delta_{\rm g}$ is the difference between the voltage gates at which $V_{\rm b}$ is larger than its values at closed gate plus $0.1 \Delta_{\rm b} $ on both sides. It can be observed that this region becomes smaller as the coupling is increased whereas its width increases with the coupling. The width also decreases steadily as the temperature of the drain increases whereas the behavior of its width with temperature is less trivial. It decreases at large couplings whereas it increases as the temperature increases at small couplings.

\section{Couplings to the leads and single-level nature of the dot}

Here we want to show how we proceeded to determine the magnitude of the tunnel coupling of the QD to the source and the drain, namely $\Gamma_{\rm s}$ and $\Gamma_{\rm d}$. Moreover we will also show how we assessed the single-level nature ruling out the possibility of the presence of more levels.
From the microscopic point of view, the coupling depends on the wave-functions of the level considered as well as the density of state of the leads at that energy. For this reason it can vary appreciably for different resonances. Here we focus on the resonance chosen in the main text to describe the heat valve effect as this is the focus of our work.
By comparing the heat flowing into the source lead shown in Fig. \ref{fig:cooling} panel b) with the temperature maps in Figure 4 of the main text, we see that the cooling region corresponds to the region where heat flows away from the source. Therefore there is a strict connection between the cooling region in the temperature map and the region where the heat flow is negative.

From Equation \ref{eq:hcur}, and assuming for now a single-level at energy $v_{\rm g}$, the heat current is given by:
\begin{equation}
\dot Q_{{\rm D}}(V_{\rm g},V_{\rm b})=\int \frac {d\omega}{2\pi}(\omega-\mu_{\rm s}) \frac{r(1+r)^{-2}\Gamma^2 }{(\omega-v_{\rm g})^2+(\Gamma_{\rm T}/2)^2} \Delta f(\omega)
\label{eq:heatcur}
\end{equation}
where $\Delta f(\omega)(f_{\rm s}(\omega)-f_{\rm d}(\omega))$ is the Fermi-Dirac distribution, $\Gamma=\Gamma_{\rm s}+\Gamma_{\rm d}$ is the total coupling to the leads and $r=\Gamma_{\rm s}/\Gamma_{\rm d}$ is the asymmetry in the coupling. The total broadening given by $\Gamma_{\rm T}=\Gamma+\Gamma_{\rm ext}$ includes some extra contribution $\Gamma_{\rm ext}$ which accounts for extra broadening mechanisms such as fluctuations of the applied gate voltage. 
The above expression tells us that the total broadening $\Gamma_{\rm T}$ is responsible for the width of the curve $\dot Q_{{\rm D}}(V_{\rm g},0)$ whereas the height of the same curve is determined by both the total coupling $\Gamma$ and the asymmetry $r=\Gamma_{\rm s}/\Gamma_{\rm d}$. It is easy to check that, at fixed $\Gamma$ and $\Gamma_{\rm ext}$ the maximum heat current is achieved at $r=1$. 
We have seen in the previous section that the width of the cooling region at $V_{\rm b}=0$ depends crucially only on the total width $\Gamma_{\rm T}$, namely the width of the spectral function of the quantum dot.

\begin{figure}[]
	\includegraphics[width=0.6\linewidth]{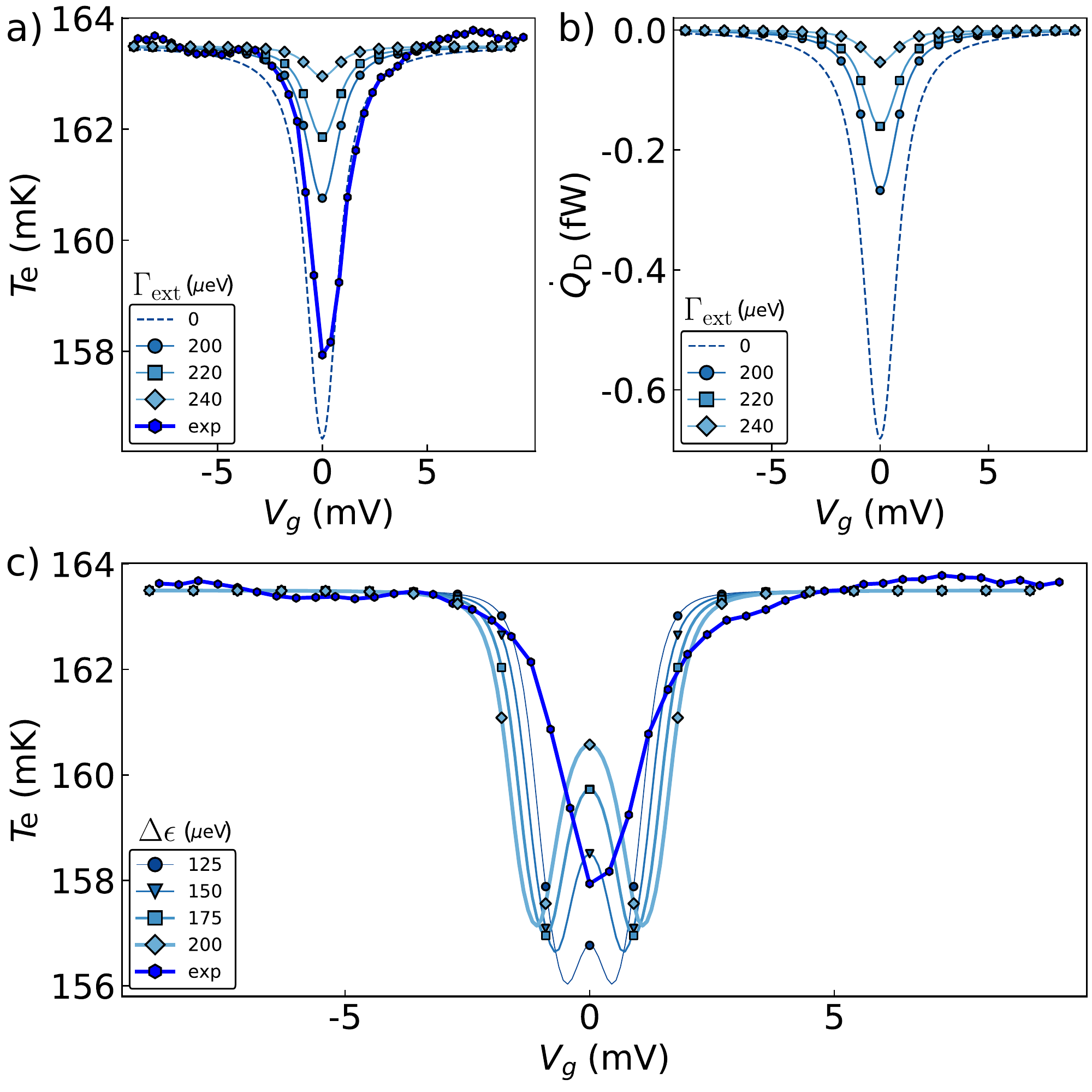}
	\caption{(Color online). a) Temperature profiles at $V_{\rm b}=0$ as a function of the gate voltage. The thick solid line is the curve obtained from the experimental measurement whereas the dashed line is the theoretical prediction obtained with the same parameters as in \mbox{Fig. 4} of the main text. The other three curves are for different values of the extra broadening $\Gamma_{\rm ext}$ and for $\Gamma_{\rm s}=\Gamma_{\rm d}$ where the maximum cooling power is obtained. b) Theoretical results for the heat current corresponding to the temperature variations in the left panel. c) Temperature curves for the case of a two-level system compared to the experimental curve at $V_{\rm b}=0$. Each curve corresponds to a different level separation $\Delta \epsilon$.}
	\label{fig:single}	
\end{figure}

Let us now assume that the broadening observed in the temperature map in \mbox{Fig. 4} of the main text is mostly due to some external source therefore having $\Gamma\ll \Gamma_{\rm ext}$. In this case the total heat current would be greatly reduced. In \mbox{Fig. \ref{fig:single}} panel a) we compare the temperature profile as a function of the gate voltage at zero bias $V_{\rm b}=0$ for (dashed) the theoretical calculation used in \mbox{Fig. 4} of the main text and (thin solid lines with markers) the case of $\Gamma_{\rm T}=250 \muev$, $r=1$ and different $\Gamma$ such that $\Gamma \ll\Gamma_{\rm ext}$. We have also added the experimentally measured temperature profile (thick solid line). In panel b) of \mbox{Fig. \ref{fig:single}} we plot the corresponding heat currents. We can see that if the broadening would be due mostly to some other mechanism other than the coupling to the leads, even in the best case scenario of symmetric coupling ($r=1$) where the cooling power is maximum, we would not reach the current needed for the observed cooling. We therefore conclude that the main mechanism for the broadening of the spectral function is due to the coupling to the leads.

The asymmetry parameter $r=3/17$ used in \mbox{Fig. 4} of the main text has been chosen in order to get the best match between the observed temperature profiles and the computed ones with the inbedding technique. We therefore conclude that the resonance used in heat valve effect discussed in the main text is strongly coupled to the leads. Furthermore we have shown that any other broadening mechanism, if present at all, does not contribute substantially to the width of the spectral function of the level.

We now move to the discussion of the single-level nature of the QD. \mbox{Figure~\ref{fig:single}} panel c) shows the theoretical prediction for the temperature variation as a function of the gate voltage for the case of two levels and compares it with the single-level prediction and the experimental data. We have used $\Gamma=0.25 \muev$ as established above and we have set the ratio $r=3/17$ for the two-level case too. The separation $\Delta \epsilon$ between the two levels has been chosen to be $\lesssim \Gamma/2$ in order to make them distinguishable from a single degenerate level but such that the two levels are not completely separated. In this latter case their effect would be the same as that of two single-level resonances. It is clear from this comparison that if more than one level is involved the temperature profile at zero bias would be markedly different.
We therefore conclude that the measured temperature profile is indeed consistent with a strongly coupled single level, which is consistent with the charge transport data.


\begin{thebibliography}{29}
\bibitem{Pop2010} E. Pop, "Energy dissipation and transport in nanoscale devices", Nano Research {\bf 3}, 147 (2010).
\bibitem{Lee2013} W. Lee, K. Kim, W. Jeong, L. Zotti, F. Pauly, J. C. Cuevas, P. Reddy, "Heat dissipation in atomic-scale junctions", Nature {\bf 498}, 209 (2013).
\bibitem{Zotti2014} L. Zotti, M. B{\"u}rkle, F. Pauly, W. Lee, K. Kim, W. Jeong, Y. Asai, P. Reddy, J. C. Cuevas, "Heat dissipation and its relation to thermopower in single-molecule junctions", New J. Phys. {\bf 16}, 015004 (2014).
\bibitem{Dubi2011} Y. Dubi and M. Di Ventra, "Heat flow and thermoelectricity in atomic and molecular junctions", Rev. Mod. Phys. {\bf 83}, 131 (2011).

\bibitem{Kouwenhoven2001} L. P. Kouwenhoven, D. G. Austing, S. Tarucha, "Few-electron quantum dots", Rep. Prog. Phys. {\bf 64}, 701 (2001).

\bibitem{Scheibner2005} R. Scheibner, H. Buhmann, D. Reuter, M. N. Kiselev, and L. W. Molenkamp, "Thermopower of a Kondo Spin-Correlated Quantum Dot",
Phys. Rev. Lett. {\bf 95}, 176602 (2005)
\bibitem{Godijn1999} S. F. Godijn, S. M\" oller, H. Buhmann, and L. W. Molenkamp, S. A. van Langen, "Thermopower of a Chaotic Quantum Dot, Phys. Rev. Lett. {\bf 82}, 2927 (1999).
\bibitem{Dzurak1997} A. S. Dzurak, C. G. Smith, C. H. W. Barnes, M. Pepper, L. Mart\~in-Moreno, C. T. Liang, D. A. Ritchie, and G. A. C. Jones, "Thermoelectric signature of the excitation spectrum of a quantum dot", Phys. Rev. B {\bf 55}, R10197 (1997).

\bibitem{Josefsson2018} M. Josefsson, A. Svilans, A. M. Burke, E. A. Hoffmann, S. Fahlvik, C. Thelander, M. Leijnse, H. Linke, "A quantum-dot heat engine operating close to the thermodynamic efficiency limits", Nature Nanotech. {\bf 13}, 1 (2018).
\bibitem{Jaliel-PRL19} G. Jaliel, R. K. Puddy, R. Sanchez, A. N. Jordan, B. Sothmann, I. Farrer, J. P. Griffiths, D. A. Ritchie, and C. G. Smith, "Experimental Realization of a Quantum Dot Energy Harvester", Phys. Rev. Lett. {\bf 123}, 117701 (2019).
\bibitem{Prete-NanoLett19} D. Prete, P.  A. Erdman, V. Demontis, V. Zannier, D. Ercolani, L. Sorba, F. Beltram, F. Rossella, F. Taddei, S. Roddaro, "Thermoelectric Conversion at 30 K in InAs/InP Nanowire Quantum Dots", Nano Lett. {\bf 19}, 3033 (2019).

\bibitem{Kim2001} P. Kim, L. Shi, A. Majumdar, and P. L. McEuen, "Thermal transport measurements of individual multiwalled nanotubes", Phys. Rev. Lett. {\bf 87}, 215502 (2001).
\bibitem{Huang2007} Z. Huang, F. Chen, R. D'Agosta, P. A. Bennett, M. Di Ventra, and N. Tao, "Local ionic and electron heating in single-molecule junctions", Nature Nanotech. {\bf 2}, 698 (2007).
\bibitem{Tsutsui2008} M. Tsutsui, M. Taniguchi, and T. Kawai, "Local Heating in Metal--Molecule--Metal Junctions", Nano Lett. {\bf 8}, 3293 (2008).

\bibitem{Halbertal2016} D. Halbertal, J. Cuppens, M. Ben Shalom, L. Embon, N. Shadmi, Y. Anahory, H. R. Naren, J. Sarkar, A. Uri, Y. Ronen, Y. Myasoedov, L. S. Levitov, E. Joselevich, A. K. Geim and E. Zeldov, "Nanoscale thermal imaging of dissipation in quantum systems", Nature {\bf 539}, 407 (2016).
\bibitem{Marguerite2019} A. Marguerite, J. Birkbeck, A. Aharon-Steinberg, D. Halbertal, K. Bagani, I. Marcus, Y. Myasoedov, A. K. Geim, D. J. Perello, and E. Zeldov, "Imaging work and dissipation in the quantum Hall state in graphene", Nature {\bf 575}, 628 (2019).
\bibitem{Molenkamp1992} L. W. Molenkamp, Th. Gravier, H. van Houten, O. J. A. Buijk, M. A. A. Mabesoone, and C. T. Foxon, "Peltier coefficient and thermal conductance of a quantum point contact", Phys. Rev. Lett. {\bf 68}, 3765 (1992).

\bibitem{Prance-PRL09} J. R. Prance, C. G. Smith, J. P. Griffiths, S. J. Chorley, D. Anderson, G. A. C. Jones, I. Farrer, and D. A. Ritchie, "Electronic Refrigeration of a Two-Dimensional Electron Gas", Phys. Rev. Lett. {\bf 102}, 146602 (2009).
\bibitem{Jezouin2013} S. J\'ezouin, F. D. Parmentier, A. Anthore, U. Gennser, A. Cavanna, Y. Jin, and F. Pierre, "Quantum Limit of Heat Flow Across a Single Electronic Channel", Science {\bf 342}, 601 (2013).
\bibitem{Sivre2018} E. Sivre, A. Anthore, F. D. Parmentier, A. Cavanna, U. Gennser, A. Ouerghi, Y. Jin, and F. Pierre, "Heat Coulomb blockade of one ballistic channel", Nature Phys. {\bf 14}, 145 (2018).
\bibitem{Banerjee2018} M. Banerjee, M. Heiblum, V. Umansky, D. E. Feldman, Y. Oreg, and A. Stern, "Observation of half-integer thermal Hall conductance", Nature {\bf 559}, 205 (2018) .
\bibitem{Tikhonov2016} E. S. Tikhonov, D V Shovkun, D Ercolani, F Rossella, M Rocci, L Sorba, S Roddaro and V S Khrapai, "Noise thermometry applied to thermoelectric measurements in InAs nanowires", Semicond. Sci. Technol. {\bf 31}, 104001 (2016).

\bibitem{Nahum1995} M. Nahum and J. M. Martinis, "Ultrasensitive hot electron microbolometer", Appl. Phys. Lett. {\bf 63}, 3075 (1993).
\bibitem{Giazotto-RMP} F. Giazotto, T. T. Heikkil\"a, A. Luukanen, A. M. Savin and J. P. Pekola, "Opportunities for mesoscopics in thermometry and refrigeration: Physics and applications", Rev. Mod. Phys. {\bf 78}, 217 (2006).

\bibitem{Meschke-JLTP09} M. Meschke, J. T. Peltonen, H. Courtois, J. P. Pekola, "Calorimetric Readout of a Superconducting Proximity-Effect Thermometer", J. Low Temp. Phys. {\bf 154}, 190 (2009).
\bibitem{WangAPL18} L. B. Wang, O. P. Saira, J.-P. Pekola, "Fast thermometry with a proximity Josephson junction", Appl. Phys. Lett.{\bf 112}, 013105 (2018).

\bibitem{Giazotto-Nature12} F. Giazotto, and M. J. Mart\~inez-Perez, "The Josephson heat interferometer", Nature {\bf 492}, 41 (2012).
\bibitem{Ronzani-NatPhys18} A. Ronzani, B. Karimi, J. Senior, C. Chen, and J. P. Pekola, "Tunable photonic heat transport in a quantum heat valve", Nature Phys. {\bf 114}, 991 (2018).

\bibitem{Saira2007} O. P. Saira, M. Meschke, F. Giazotto, A. M. Savin, M. M\" ott\" onen, J. P. Pekola, "Heat transistor: demonstration of gate-controlled electronic refrigeration", Phys. Rev. Lett. {\bf 99}, 027203 (2007).

\bibitem{Dutta-PRL17} B. Dutta, J. T. Peltonen, D. S. Antonenko, M. Meschke, M. A. Skvortsov, B. Kubala, J. K\"onig, C. B. Winkelmann, H. Courtois, and J. P. Pekola, "Thermal Conductance of a Single-Electron Transistor", Phys. Rev. Lett. {\bf 119}, 077701 (2017).
\bibitem{Kubala-PRL08}  B. Kubala, J. K\" onig, and J. Pekola, "Violation of the Wiedemann-Franz Law in a Single-Electron Transistor", Phys. Rev. Lett. {\bf 100}, 066801 (2008).

\bibitem{Edwards-APL93} H. L. Edwards, Q. Niu, A. L. de Lozanne, "A quantum dot refrigerator", Appl. Phys. Lett. {\bf 63}, 1815 (1993).

\bibitem{Dutta-Nanolett19} B. Dutta, D. Majidi, A. Garcia Corral, P. A. Erdman, S. Florens, T. A. Costi, H. Courtois, C. B. Winkelmann, "Direct probe of the Seebeck coefficient in a Kondo-correlated single-quantum-dot transistor", Nano Letters {\bf 19}, 506 (2019).

\bibitem{Kontos2004} T. Kontos, M. Aprili, J. Lesueur, X. Grison, and L. Dumoulin, "Superconducting Proximity Effect at the Paramagnetic-Ferromagnetic Transition", Phys. Rev. Lett. {\bf 93}, 137001 (2004).
\bibitem{Park-APL99} H. Park, A. K. Lim,  A. P. Alivisatos, J. Park, and P. L. McEuen, "Fabrication of metallic electrodes with nanometer separation by electromigration", Appl. Phys. Lett. {\bf 75}, 301 (1999).


\bibitem{ChargingEnergy} Here we use the definition $E_{\rm c}=e^2/2C_{\rm \Sigma}$ where $C_{\rm \Sigma}$ is the total capacitance of the dot with respect to leads and gate.
\bibitem{suppmat} See Supplemental Material, which includes Refs. [38-43].
\bibitem{Bolotin2004} K. I. Bolotin, F. Kuemmeth, A. N. Pasupathy, and D. C. Ralph, "Metal-nanoparticle single-electron transistors fabricated using electromigration", Appl. Phys. Lett. {\bf 84}, 3154 (2004).
\bibitem{Ralph-PRL95} D. C. Ralph, C. T. Black, and M. Tinkham, "Spectroscopic Measurements of Discrete Electronic States in Single Metal Particles", Phys. Rev. Lett. {\bf 74}, 3241 (1995).
\bibitem{vanZanten-PRB16} D. M. T. van Zanten, F. Balestro, H. Courtois, and C. B. Winkelmann, "Probing hybridization of a single energy level coupled to superconducting leads", Phys. Rev B {\bf 92}, 184501 (2015).
\bibitem{Thijssen-PSSB08} J. M. Thijssen and H. S. J. van der Zant, "Charge transport and single-electron effects in nanoscale systems", Phys. Stat. Sol. (b) {\bf 245}, 1455 (2008)
\bibitem{Bonet-2002} E. Bonet, M. M. Deshmukh, and D. C. Ralph, "Solving rate equations for electron tunneling via discrete quantum states", Phys. Rev. B {\bf 65}, 045317 (2002).
\bibitem{Joonas-PRL2010} J. T. Peltonen, P. Virtanen, M. Meschke, J. V. Koski, T. T. Heikkil\"a, J. P. Pekola, "Thermal conductance by the inverse proximity effect in a superconductor", Phys. Rev. Lett. {\bf 105}, 097004 (2010).


\bibitem{Dubos-PRB01} P. Dubos, H. Courtois, B. Pannetier, F. K. Wilhelm, A. D. Zaikin, and G. Sch\" on, "Josephson critical current in a long mesoscopic S-N-S junction", Phys. Rev. B {\bf 63}, 064502 (2001).

\bibitem{stefleebooka} G. Stefanucci and R. van Leeuwen, "Nonequilibrium many-body theory of quantum systems", Cambridge University Press, Cambridge, UK, 2013.

\bibitem{Talarico-PRB20} N. W. Talarico, S. Maniscalco, N. Lo Gullo, "Study of the energy variation in many-body open quantum systems: Role of interactions in the weak and strong coupling regimes", Phys. Rev. B {\bf  101}, 045103 (2020).

\bibitem{Harzheim-PRR20} A. Harzheim, J. K. Sowa, J. L. Swett, G. A. D. Briggs, J. A. Mol, and P. Gehring, "Role of metallic leads and electronic degeneracies in thermoelectric power generation in quantum dots", Phys. Rev. Res. {\bf 2}, 013140 (2020).
\bibitem{RevuePekola} J. P. Pekola, "Towards quantum thermodynamics in electronic circuits", Nature Physics {\bf 11}, 118 (2015).

\end{thebibliography}

\begin{thebibliography}{99}
\bibitem{Bolotin2004} K. I. Bolotin, F. Kuemmeth, A. N. Pasupathy, and D. C. Ralph, "Metal-nanoparticle single-electron transistors fabricated using electromigration", Appl. Phys. Lett. {\bf 84}, 3154 (2004).
\bibitem{Dutta-Nanolett19} B. Dutta, D. Majidi, A. Garcia Corral, P. A. Erdman, S. Florens, T. A. Costi, H. Courtois, C. B. Winkelmann, "Direct probe of the Seebeck coefficient in a Kondo-correlated single-quantum-dot transistor", Nano Letters {\bf 19}, 506 (2019).
\bibitem{Ralph-PRL95} D. C. Ralph, C. T. Black, and M. Tinkham, "Spectroscopic Measurements of Discrete Electronic States in Single Metal Particles", Phys. Rev. Lett. {\bf 74}, 3241 (1995).
\bibitem{vanZanten-PRB16} D. M. T. van Zanten, F. Balestro, H. Courtois, and C. B. Winkelmann, "Probing hybridization of a single energy level coupled to superconducting leads", Phys. Rev B {\bf 92}, 184501 (2015).
\bibitem{Thijssen-PSSB08} J. M. Thijssen and H. S. J. van der Zant, "Charge transport and single-electron effects in nanoscale systems", Phys. Stat. Sol. (b) {\bf 245}, 1455 (2008)
\bibitem{Bonet-2002} E. Bonet, M. M. Deshmukh, and D. C. Ralph, "Solving rate equations for electron tunneling via discrete quantum states", Phys. Rev. B {\bf 65}, 045317 (2002).
\bibitem{Dubos-PRB01b} P. Dubos, H. Courtois, B. Pannetier, F. K. Wilhelm, A. D. Zaikin, and G. Sch\" on, "Josephson critical current in a long mesoscopic S-N-S junction", Phys. Rev. B {\bf 63}, 064502 (2001).
\bibitem{Joonas-PRL2010} J. T. Peltonen, P. Virtanen, M. Meschke, J. V. Koski, T. T. Heikkil\"a, J. P. Pekola, "Thermal conductance by the inverse proximity effect in a superconductor", Phys. Rev. Lett. {\bf 105}, 097004 (2010).
\bibitem{Giazotto-RMP06} F. Giazotto, T. T. Heikkil\"a, A. Luukanen, A. M. Savin and J. P. Pekola, "Opportunities for mesoscopics in thermometry and refrigeration: Physics and applications", Rev. Mod. Phys. {\bf 78}, 217 (2006).
\bibitem{stefleebook} G. Stefanucci and R. van Leeuwen, "Nonequilibrium many-body theory of quantum systems", Cambridge University Press, Cambridge, UK, 2013.
\end{thebibliography}
\end{document}